# A method for command identification, using modified collision free hashing with addition & rotation iterative hash functions (part 1)*.

Dimitrios Skraparlis

Aristotle University of Thessaloniki, Department of Electrical & Computer Engineering

e-mail: kounoupidas@hotmail.com

**Abstract.** This paper proposes a method for identification of a user's fixed string set (which can be a command/instruction set for a terminal or microprocessor). This method is fast and has very small memory requirements, compared to a traditional full string storage and compare method. The user feeds characters into a microcontroller via a keyboard or another microprocessor sends commands and the microcontroller hashes the input in order to identify valid commands, ensuring no collisions between hashed valid strings, while applying further criteria to narrow collision between random and valid strings. The method proposed narrows the possibility of the latter kind of collision, achieving small code and memory-size utilization and very fast execution. Hashing is achieved using additive & rotating hash functions in an iterative form, which can be very easily implemented in simple microcontrollers and microprocessors. Such hash functions are presented and compared according to their efficiency for a given string/command set, using the program found in the appendix.

## Introduction

What is hashing? Hash functions operate on a string of arbitrary length and output a short string of fixed length. For example, the string „perform a reset" (of length 15) could be hashed using instructions that operate on a block size of 2 characters, resulting in an output like „oy", which has a length of 2 (86.66% efficiency). The hash functions analyzed in this paper have a hash length equal to the block size, like in the example above. The block size is very small (1 byte or perhaps 1 word), allowing only relatively small strings sets (ideally, for a block size of 1 byte, a maximum string set of 256 unique commands/strings is allowed, though presenting a high collision coincidence between hashed valid and hashed invalid commands).

### The need for such functions

How can a microcontroller identify arbitrary-length commands sent to it? The simplest solution would be to store the full strings representing the commands and have a simple algorithm that compares the input to this string table. Such a solution was considered impractical (especially for a terminal application) because the string/command set would be limited by the microcontroller's small memory; each string would require memory equal to its length in order for it to be stored.

Why not use hash functions? One could store only the hashed values instead of storing the whole string. The hash length could be much smaller than the strings' length and we would expect for example a decrease of memory usage reaching 91.66% for strings of length 12, using a hash length of one.

### Definitions

The functions used in this paper are simple additive & rotating hash functions: they use only some exclusive-or, add and shift left/right instructions commonly found in general purpose microcontrollers. No modulos or prime numbers are used.

Collisions occur when the results of a hash function operating on two different inputs are the same:
*Intrinsic collision* is the collision between two hashed valid strings.
*Extrinsic collision* is the collision between a hashed valid and a hashed invalid string.

---

* Part 2 of this paper will be published in the future, as a complete software solution in C++ is being implemented by the author in order to fully study the characteristics of the suggested hash functions.

There is also another kind of collision, which occurs when the „same length" criterion is applied and certain interface behaviour has been chosen (it is assumed that the protocol does not accept more characters at the input if the value and the length of the input string are those of a valid string/command or that the commands are executed once their valid hashed values and length are detected): if an invalid string (let X) hashes to the same value as a valid string (let A) and they also have the same length, there is a possibility that X would be the first part of another valid string (let B), which, in that case, cannot be recognized (A will be identified instead of B). As an example, let one consider that:

„go" hashes to 9,
„st" hashes to 9 and
„stop" hashes to 7.

„go" and „stop" are valid commands, while „st" is not. However, if „st" is the input, „go" will be identified and further input will be recognized as a new string.

## The hash functions

### Suggested hash functions

Possible iterated addition & rotation hash functions are presented in the following table. Their performance (given a fixed string set) is then analysed. The target of this analysis is not only to assure collision-free identification of each string but also to evaluate the performance of each hash function (the collision percentage) during a brute-force hashing of random strings. To do this, a program was written – the code is found in the appendix.

The following iterative hash functions are generated by the general iterative function

$$M_i = M_i \oplus (f(H_{i-1}) \oplus g(M_{i-1}))$$

where f, g are some functions(simple addition or rotation instructions) operating on their inputs.

| Iterative hash function | Comments |
|---|---|
| $H = M \oplus (H + 0xAA) \oplus (M>>1)$ | $0xAA = (AA)_{16} = (10101010)_2$ |
| $H = M \oplus (H + 0x55) \oplus (M<<1)$ | $0x55 = (55)_{16} = (01010101)_2$ |
| $H = M \oplus (M + 0xAA) \oplus (H>>1)$ | same as above |
| $H = M \oplus (M + 0x55) \oplus (H<<1)$ | same as above |
| $H = M \oplus (M + length - x) \oplus (H>>x)$ | |
| $H = M \oplus (M + length - x) \oplus (H<<x)$ | |
| $H = M \oplus (M>>x) \oplus (H<<x)$ | Characters contribute more to the final result as more as their position in the string approaches the first |
| $H = M \oplus (M<<x) \oplus (H>>x)$ | Characters contribute more to the final result as more as their position in the string approaches the first |
| $H = M \oplus H$ | Simple xor, not suggested – but can correct permutations („run" and „rnu" hash to the same value |
| $H = (M>>1) \oplus (H<<1)$ | |
| $H = (M<<1) \oplus (H>>1)$ | |
| $H = M \oplus (H \& M)$ | probably suitable only for small string lengths |
| $H = M \oplus (M>>1) \oplus (H+1)$ | |
| $H = M \oplus (M>>2) \oplus (H+1)$ | |
| $H = M \oplus (M>>y) \oplus (H+2)$ | |
| $H = M \oplus (M>>y) \oplus (H*2)$ | |
| $H = M \oplus ((H+7) \oplus (M<<1)>>1)$ | |

key:
*x* is the position of the character in the string  *H* represents the hash result
*length* is the length of the string  *M* represents the input block
*y* is any number between 1 and 5

The efficiency of each function should be recalculated for every different string set. <u>No intrinsic collisions can be tolerated, unless they refer to similar commands</u> (for example, «set speed» and «change speed» have the same effect on the system).

Some words about the symbolisation of the variables:
H appearing before the equality sign «=» is the same like $H_i$ and means the result of the hash function at the iterative step i. H appearing after the equality sign is the same like $H_{i-1}$ which indicate the result of the hash function after the previous iterative step (i-1). M is the input at the current iterative step (i). This is the main idea behind the usage of the iterative hash functions; no memory is required during the decoding of the strings – the hash function operates on the current input-block only, using the result from the previous step. In some cases, a counter could also be used in order to apply the length criterion presented below and also to influence the hash function result.

### The idea put into practice

The method is easy to implement. However, the very small hash length does not always assure collision free hashing, especially between valid and invalid strings. Three solutions exist: either one increases the hash length, applies further criteria to assure that there is no collision or uses another hash function.

The hash functions presented here can ensure intrinsic collision free hashing for the given string set, provided that one selects and uses the appropriate hash function.

The other criteria that must be met in order to assure robustness of the system against random-input feed (caused by errors at the transmission line or by the user) are presented below. A typical example of a problem that can be solved by applying these criteria is a user typing a random text of perhaps very small length (i.e. 1 character), which has the same hash result with a normal command.
It was therefore considered wise to apply further criteria that ensure a narrow collision probability:
Collision free hashing of any random string is ensured provided that
   a) the string does not have the same length as the original string and (optional)
   b) the random string does not start with the same letter as the original string or
   c) the random string does not end with the same one or two letters as the original string

The collision requirements of simple terminal applications are satisfied by meeting the first condition only, as the possibility of extrinsic collision is too low. Please refer to the evaluation part of this paper.

It should also be noted that a single hash function suitable for all users' string sets will not be suggested here – additive & rotating hash function results depend on the application (the fixed string set). The user is responsible for selecting the appropriate hash functions that ensures collision free hashing between valid strings (no intrinsic collisions) and (if desired) a low possibility of hash values occurance between any valid and invalid strings (as less number of extrinsic collisions as possible).

### Evaluation

The following results were generated using the program found in the appendix. The command string set is for a terminal that controls the operation of a motor. Note that the command/instruction strings are converted to strings that can be directly printed to the Liquid Crystal Display used (according to the LCD controller's font set). For example, the ASCII code for „a" is 97, while the controller's font set C code for „a" is 225 (=97+128). For this reason, a kind of offset was used inside the program found in the appendix (offset equals 128). Using the same hash function on the ASCII strings will of course produce different results.

Evaluation for many addition & rotation hash functions follows:

A method for command identification, using modified collision free hashing with addition & rotation iterative hash functions. (Part 1)

**evaluation of the iterative hash function: H = M Xor H**

evaluate hash results | find random strings with the same length

[hash the string set] | Count coincidences | Text that hashes to the same value | Text that hashes to the same value and has the same starting character | Text that hashes to the same value and has the same ending character | Text that hashes to the same value and has the same 2 ending characters

number of coincidences: **0**

coincidence table:

efficiency and classification: Maximum coincidence count is 1 with a hash coincidence efficiency of 100%. The hash function is classified as EXCELLENT.

| | string set | length | hashed | number of hits / possibility of occurance % (P): | | | |
|---|---|---|---|---|---|---|---|
| ☑ | info | 4 | 14 | 14239 /P=3,115919% | 549 /P=0,1201376% | 543 /P=0,1188246% | 21 /P=4,595427E-03% |
| ☑ | start | 5 | 224 | 371039 /P=3,122862% | 14239 /P=0,119843% | 14239 /P=0,119843% | 549 /P=4,620677E-03% |
| ☑ | stop | 4 | 24 | 14287 /P=3,126422% | 555 /P=0,1214506% | 555 /P=0,1214506% | 21 /P=4,595427E-03% |
| ☑ | reset | 5 | 245 | 371535 /P=3,127037% | 14319 /P=0,1205163% | 14351 /P=0,1207857% | 549 /P=4,620677E-03% |
| ☑ | help | 4 | 17 | 14239 /P=3,115919% | 543 /P=0,1188246% | 543 /P=0,1188246% | 19 /P=4,157768E-03% |
| ☐ | destspeed | 9 | 225 | | | | |
| ☐ | setspeed | 8 | 5 | | | | |
| ☐ | newspeed | 8 | 27 | | | | |
| ☐ | targetspeed | 11 | 246 | | | | |
| ☑ | ds | 2 | 23 | 19 /P=2,810651% | 0 /P=0% | 0 /P=0% | -1 /P=-0,147929% |
| ☑ | ss | 2 | 0 | 25 /P=3,698225% | 0 /P=0% | 0 /P=0% | -1 /P=-0,147929% |
| ☑ | ns | 2 | 29 | 21 /P=3,106509% | 0 /P=0% | 0 /P=0% | -1 /P=-0,147929% |
| ☑ | ts | 2 | 7 | 21 /P=3,106509% | 0 /P=0% | 0 /P=0% | -1 /P=-0,147929% |
| ☐ | turnleft | 8 | 6 | | | | |
| ☐ | turnright | 9 | 253 | | | | |
| ☑ | max | 3 | 244 | 555 /P=3,157715% | 19 /P=0,108102% | 19 /P=0,108102% | 0 /P=0% |
| ☑ | min | 3 | 234 | 555 /P=3,157715% | 21 /P=0,1194811% | 21 /P=0,1194811% | 0 /P=0% |
| ☑ | tcnt | 4 | 13 | 14239 /P=3,115919% | 543 /P=0,1188246% | 543 /P=0,1188246% | 19 /P=4,157768E-03% |
| ☐ | interval | 8 | 31 | | | | |

☐ Show results

**evaluation of the iterative hash function: H = M Xor (M / 2) Xor (H * 2)**

evaluate hash results | find random strings with the same length

[hash the string set] | Count coincidences | Text that hashes to the same value | Text that hashes to the same value and has the same starting character | Text that hashes to the same value and has the same ending character | Text that hashes to the same value and has the same 2 ending characters

number of coincidences: **2**

coincidence table: #180:2

efficiency and classification: Maximum coincidence count is 2 with a hash coincidence efficiency of 89,4736842105263%. The hash function is classified as BAD / INAPPROPRIATE.

| | string set | length | hashed | number of hits / possibility of occurance % (P): | | | |
|---|---|---|---|---|---|---|---|
| ☑ | info | 4 | 49 | 3112 /P=0,6809986% | 249 /P=5,448864E-02% | 235 /P=5,142502E-02% | 16 /P=3,501278E-03% |
| ☑ | start | 5 | 60 | 20987 /P=0,1766378% | 758 /P=6,379732E-03% | 5357 /P=4,508737E-02% | 413 /P=3,476028E-03% |
| ☑ | stop | 4 | 214 | 656 /P=0,1435524% | 54 /P=1,181681E-02% | 169 /P=3,698225E-02% | 2 /P=4,376597E-04% |
| ☑ | reset | 5 | 140 | 21743 /P=0,1830007% | 935 /P=7,869459E-03% | 5597 /P=4,710734E-02% | 429 /P=3,610693E-03% |
| ☑ | help | 4 | 0 | 933 /P=0,2041683% | 69 /P=1,509926E-02% | 73 /P=1,597458E-02% | 16 /P=3,501278E-03% |
| ☐ | destspeed | 9 | 52 | | | | |
| ☐ | setspeed | 8 | 180 | | | | |
| ☐ | newspeed | 8 | 20 | | | | |
| ☐ | targetspeed | 11 | 180 | | | | |
| ☑ | ds | 2 | 165 | 19 /P=2,810651% | 1 /P=0,147929% | 0 /P=0% | -1 /P=-0,147929% |
| ☑ | ss | 2 | 155 | 14 /P=2,071006% | 1 /P=0,147929% | 1 /P=0,147929% | -1 /P=-0,147929% |
| ☑ | ns | 2 | 187 | 16 /P=2,366864% | 1 /P=0,147929% | 0 /P=0% | -1 /P=-0,147929% |
| ☑ | ts | 2 | 149 | 11 /P=1,627219% | 1 /P=0,147929% | 0 /P=0% | -1 /P=-0,147929% |
| ☐ | turnleft | 8 | 24 | | | | |
| ☐ | turnright | 9 | 194 | | | | |
| ☑ | max | 3 | 202 | 74 /P=0,4210287% | 5 /P=2,844788E-02% | 19 /P=0,108102% | 0 /P=0% |
| ☑ | min | 3 | 207 | 239 /P=1,359809% | 16 /P=9,103323E-02% | 19 /P=0,108102% | 0 /P=0% |
| ☑ | tcnt | 4 | 136 | 727 /P=0,1590893% | 70 /P=1,531809E-02% | 180 /P=3,938938E-02% | 13 /P=2,844788E-03% |
| ☐ | interval | 8 | 164 | | | | |

☐ Show results

**A method for command identification, using modified collision free hashing with addition & rotation iterative hash functions. (Part 1)**

## evaluation of the iterative hash function: H = M Xor (M / 2) Xor (H + 2)

evaluate hash results — find random strings with the same length

[hash the string set] | Count coincidences | Text that hashes to the same value | Text that hashes to the same value and has the same starting character | Text that hashes to the same value and has the same ending character | Text that hashes to the same value and has the same 2 ending characters

number of coincidences: **4**

coincidence table:
#31:2
#145:2

efficiency and classification:
Maximum coincidence count is 2 with a hash coincidence efficiency of 78,94736842 10526%. The hash function is classified as BAD / INAPPROPRIATE.

Show results

| ☑ | string set | length | hashed | hits / P (same value) | same starting char | same ending char | same 2 ending chars |
|---|---|---|---|---|---|---|---|
| ☑ | info | 4 | 38 | 2983 /P=0,6527695% | 555 /P=0,1214506% | 113 /P=2,472777E-02% | 1 /P=2,188299E-04% |
| ☑ | start | 5 | 145 | 274400 /P=2,309497% | 12328 /P=0,103759% | 9475 /P=7,974666E-02% | 413 /P=3,476028E-03% |
| ☑ | stop | 4 | 28 | 11301 /P=2,472996% | 520 /P=0,1137915% | 378 /P=8,271769E-02% | 14 /P=3,063618E-03% |
| ☑ | reset | 5 | 158 | 267505 /P=2,251465% | 11651 /P=9,806103E-02 | 11753 /P=9,891952E-02 | 515 /P=4,334515E-03% |
| ☑ | help | 4 | 45 | 2617 /P=0,5726778% | 604 /P=0,1321732% | 85 /P=1,860054E-02% | 3 /P=6,564896E-04% |
| ☐ | destspeed | 9 | 175 | | | | |
| ☐ | setspeed | 8 | 31 | | | | |
| ☐ | newspeed | 8 | 26 | | | | |
| ☐ | targetspeed | 11 | 203 | | | | |
| ☑ | ds | 2 | 31 | 13 /P=1,923077% | 1 /P=0,147929% | 0 /P=0% | -1 /P=-0,147929% |
| ☑ | ss | 2 | 4 | 27 /P=3,994083% | 1 /P=0,147929% | 1 /P=0,147929% | -1 /P=-0,147929% |
| ☑ | ns | 2 | 20 | 20 /P=2,95858% | 1 /P=0,147929% | 0 /P=0% | -1 /P=-0,147929% |
| ☑ | ts | 2 | 7 | 13 /P=1,923077% | 1 /P=0,147929% | 0 /P=0% | -1 /P=-0,147929% |
| ☐ | turnleft | 8 | 17 | | | | |
| ☐ | turnright | 9 | 186 | | | | |
| ☑ | max | 3 | 136 | 349 /P=1,985662% | 13 /P=0,0739645% | 21 /P=0,1194811% | 0 /P=0% |
| ☑ | min | 3 | 145 | 527 /P=2,998407% | 21 /P=0,1194811% | 28 /P=0,1593082% | 0 /P=0% |
| ☑ | tcnt | 4 | 52 | 3017 /P=0,6602097% | 45 /P=9,847344E-03% | 60 /P=1,312979E-02% | 13 /P=2,844788E-03% |
| ☐ | interval | 8 | 32 | | | | |

## evaluation of the iterative hash function: H = M Xor (M / 4) Xor (H + 1)

evaluate hash results — find random strings with the same length

[hash the string set] | Count coincidences | Text that hashes to the same value | Text that hashes to the same value and has the same starting character | Text that hashes to the same value and has the same ending character | Text that hashes to the same value and has the same 2 ending characters

number of coincidences: **6**

coincidence table:
#19:2 #30:2
#222:2

efficiency and classification:
Maximum coincidence count is 2 with a hash coincidence efficiency of 68,42105263 15789%. The hash function is classified as BAD / INAPPROPRIATE.

Show results

| ☑ | string set | length | hashed | hits / P (same value) | same starting char | same ending char | same 2 ending chars |
|---|---|---|---|---|---|---|---|
| ☑ | info | 4 | 12 | 12888 /P=2,820279% | 433 /P=9,475333E-02% | 511 /P=0,1118221% | 26 /P=5,689577E-03% |
| ☑ | start | 5 | 194 | 358367 /P=3,016208% | 13849 /P=0,1165606% | 14542 /P=0,1223932% | 533 /P=4,486012E-03% |
| ☑ | stop | 4 | 6 | 12932 /P=2,829908% | 505 /P=0,1105091% | 545 /P=0,1192623% | 20 /P=4,376597E-03% |
| ☑ | reset | 5 | 208 | 318681 /P=2,682189% | 12929 /P=0,1088174% | 13554 /P=0,1140777% | 533 /P=4,486012E-03% |
| ☑ | help | 4 | 19 | 12379 /P=2,708895% | 529 /P=0,115761% | 511 /P=0,1118221% | 14 /P=3,063618E-03% |
| ☐ | destspeed | 9 | 222 | | | | |
| ☐ | setspeed | 8 | 30 | | | | |
| ☐ | newspeed | 8 | 2 | | | | |
| ☐ | targetspeed | 11 | 248 | | | | |
| ☑ | ds | 2 | 19 | 16 /P=2,366864% | 1 /P=0,147929% | 1 /P=0,147929% | -1 /P=-0,147929% |
| ☑ | ss | 2 | 30 | 19 /P=2,810651% | 1 /P=0,147929% | 1 /P=0,147929% | -1 /P=-0,147929% |
| ☑ | ns | 2 | 26 | 33 /P=4,881657% | 1 /P=0,147929% | 1 /P=0,147929% | -1 /P=-0,147929% |
| ☑ | ts | 2 | 7 | 21 /P=3,106509% | 1 /P=0,147929% | 1 /P=0,147929% | -1 /P=-0,147929% |
| ☐ | turnleft | 8 | 51 | | | | |
| ☐ | turnright | 9 | 253 | | | | |
| ☑ | max | 3 | 196 | 448 /P=2,54893% | 17 /P=0,0967228% | 26 /P=0,147929% | 0 /P=0% |
| ☑ | min | 3 | 222 | 511 /P=2,907374% | 21 /P=0,1194811% | 14 /P=7,965408E-02% | 0 /P=0% |
| ☑ | tcnt | 4 | 14 | 13426 /P=2,93801% | 572 /P=0,1251707% | 621 /P=0,1358933% | 16 /P=3,501278E-03% |
| ☐ | interval | 8 | 38 | | | | |

**A method for command identification, using modified collision free hashing with addition & rotation iterative hash functions. (Part 1)**

## evaluation of the iterative hash function: H = M Xor M / 2 Xor (H + 1)

[hash the string set]   evaluate hash results   find random strings with the same length

| ☒ | string set | length | hashed | Count coincidences | Text that hashes to the same value | Text that hashes to the same value and has the same starting character | Text that hashes to the same value and has the same ending character | Text that hashes to the same value and has the same 2 ending characters |
|---|---|---|---|---|---|---|---|---|
| | | | | number of coincidences: 6 | number of hits / possibility of occurance % (P): | | | |
| ☑ | info | 4 | 6 | | 15320 /P=3,352474% | 607 /P=0,1328297% | 657 /P=0,1437712% | 28 /P=6,127236E-03% |
| ☑ | start | 5 | 156 | | 362640 /P=3,052172% | 14296 /P=0,1203228% | 11948 /P=0,1005607% | 457 /P=3,846356E-03% |
| ☑ | stop | 4 | 40 | coincidence table: #16:2 #18:2 #25:2 | 644 /P=0,1409264% | 31 /P=6,783726E-03% | 464 /P=0,1015371% | 13 /P=2,844788E-03% |
| ☑ | reset | 5 | 141 | | 322386 /P=2,713373% | 12189 /P=0,1025891% | 15465 /P=0,1301617% | 643 /P=5,411831E-03% |
| ☑ | help | 4 | 25 | | 12227 /P=2,675633% | 583 /P=0,1275778% | 657 /P=0,1437712% | 12 /P=2,625959E-03% |
| ☐ | destspeed | 9 | 128 | | | | | |
| ☐ | setspeed | 8 | 3 | | | | | |
| ☐ | newspeed | 8 | 16 | | | | | |
| ☐ | targetspeed | 11 | 148 | | | | | |
| ☑ | ds | 2 | 17 | efficiency and classification: | 13 /P=1,923077% | 1 /P=0,147929% | 0 /P=0% | -1 /P=-0,147929% |
| ☑ | ss | 2 | 0 | Maximum coincidence count is 2 with a hash coincidence efficiency of 68,42105263 15789%. The hash function is classified as BAD / INAPPROPRIATE. | 32 /P=4,733728% | 1 /P=0,147929% | 1 /P=0,147929% | -1 /P=-0,147929% |
| ☑ | ns | 2 | 16 | | 23 /P=3,402367% | 1 /P=0,147929% | 0 /P=0% | -1 /P=-0,147929% |
| ☑ | ts | 2 | 25 | | 14 /P=2,071006% | 1 /P=0,147929% | 0 /P=0% | -1 /P=-0,147929% |
| ☐ | turnleft | 8 | 33 | | | | | |
| ☐ | turnright | 9 | 171 | | | | | |
| ☑ | max | 3 | 143 | | 442 /P=2,514793% | 11 /P=6,258535E-02% | 23 /P=0,1308603% | 0 /P=0% |
| ☑ | min | 3 | 158 | | 645 /P=3,669777% | 28 /P=0,1593082% | 27 /P=0,1536186% | 0 /P=0% |
| ☑ | tcnt | 4 | 18 | | 15115 /P=3,307614% | 437 /P=9,562866E-02% | 441 /P=9,650397E-02% | 17 /P=3,720108E-03% |
| ☐ | interval | 8 | 18 | ☐ Show results | | | | |

## evaluation of the iterative hash function: H = M Xor (H And M)

[hash the string set]   evaluate hash results   find random strings with the same length

| ☒ | string set | length | hashed | Count coincidences | Text that hashes to the same value | Text that hashes to the same value and has the same starting character | Text that hashes to the same value and has the same ending character | Text that hashes to the same value and has the same 2 ending characters |
|---|---|---|---|---|---|---|---|---|
| | | | | number of coincidences: 8 | number of hits / possibility of occurance % (P): | | | |
| ☑ | info | 4 | 15 | | 2462 /P=0,5387592% | 87 /P=0,0190382% | 2462 /P=0,5387592% | 41 /P=8,972025E-03% |
| ☑ | start | 5 | 228 | | 872008 /P=7,339284% | 28464 /P=0,2395682% | 102132 /P=0,8595974% | 11954 /P=0,1006112% |
| ☑ | stop | 4 | 16 | coincidence table: #4:2 #16:3 #228:3 | 39562 /P=8,657348% | 1401 /P=0,3065807% | 11954 /P=2,615892% | 675 /P=0,1477102% |
| ☑ | reset | 5 | 240 | | 930258 /P=7,829548% | 39374 /P=0,3313926% | 108387 /P=0,9122428% | 11479 /P=9,661339E-02% |
| ☑ | help | 4 | 16 | | 39562 /P=8,657348% | 1385 /P=0,3030794% | 11954 /P=2,615892% | 675 /P=0,1477102% |
| ☐ | destspeed | 9 | 228 | | | | | |
| ☐ | setspeed | 8 | 4 | | | | | |
| ☐ | newspeed | 8 | 4 | | | | | |
| ☐ | targetspeed | 11 | 228 | | | | | |
| ☑ | ds | 2 | 19 | efficiency and classification: | 4 /P=0,591716% | 1 /P=0,147929% | 2 /P=0,295858% | -1 /P=-0,147929% |
| ☑ | ss | 2 | 0 | Maximum coincidence count is 3 with a hash coincidence efficiency of 57,89473684 21053%. The hash function is classified as BAD / INAPPROPRIATE. | 127 /P=18,78698% | 6 /P=0,887574% | 1 /P=0,147929% | -1 /P=-0,147929% |
| ☑ | ns | 2 | 17 | | 20 /P=2,95858% | 4 /P=0,591716% | 3 /P=0,443787% | -1 /P=-0,147929% |
| ☑ | ts | 2 | 3 | | 16 /P=2,366864% | 3 /P=0,443787% | 2 /P=0,295858% | -1 /P=-0,147929% |
| ☐ | turnleft | 8 | 20 | | | | | |
| ☐ | turnright | 9 | 244 | | | | | |
| ☑ | max | 3 | 248 | | 570 /P=3,243059% | 22 /P=0,1251707% | 366 /P=2,082385% | 25 /P=0,1422394% |
| ☑ | min | 3 | 238 | | 348 /P=1,979973% | 12 /P=6,827492E-02% | 272 /P=1,547565% | 10 /P=5,689577E-02% |
| ☑ | tcnt | 4 | 16 | | 39562 /P=8,657348% | 1121 /P=0,2453083% | 4611 /P=1,009025% | 507 /P=0,1109467% |
| ☐ | interval | 8 | 12 | ☐ Show results | | | | |

**A method for command identification, using modified collision free hashing with addition & rotation iterative hash functions. (Part 1)**

## evaluation of the iterative hash function: H = M Xor (M + 85) Xor (H * 2)

evaluate hash results — find random strings with the same length

| ☒ | string set | length | hashed | | Text that hashes to the same value | Text that hashes to the same value and has the same starting character | Text that hashes to the same value and has the same ending character | Text that hashes to the same value and has the same 2 ending characters |
|---|---|---|---|---|---|---|---|---|
| | | | | number of coincidences | number of hits / possibility of occurance % (P): | | | |
| ☑ | info | 4 | 29 | **5** | 3634 /P=0,7952278% | 142 /P=3,107384E-02% | 273 /P=5,974055E-02% | 35 /P=7,659046E-03% |
| ☑ | start | 5 | 211 | | 93986 /P=0,7910363% | 3607 /P=3,035844E-02% | 7053 /P=5,936181E-02% | 560 /P=4,713259E-03% |
| ☑ | stop | 4 | 207 | coincidence table: #207:2 #231:3 | 3765 /P=0,8238945% | 177 /P=3,873289E-02% | 309 /P=6,761843E-02% | 22 /P=4,814257E-03% |
| ☑ | reset | 5 | 71 | | 92225 /P=0,7762148% | 3643 /P=3,066143E-02% | 7240 /P=0,0609357% | 536 /P=4,511262E-03% |
| ☑ | help | 4 | 59 | | 3223 /P=0,7052887% | 99 /P=2,166416E-02% | 197 /P=4,310948E-02% | 18 /P=3,938938E-03% |
| ☐ | destspeed | 9 | 231 | | | | | |
| ☐ | setspeed | 8 | 231 | | | | | |
| ☐ | newspeed | 8 | 39 | | | | | |
| ☐ | targetspeed | 11 | 231 | | | | | |
| ☑ | ds | 2 | 1 | efficiency and classification: | 16 /P=2,366864% | 1 /P=0,147929% | 1 /P=0,147929% | -1 /P=-0,147929% |
| ☑ | ss | 2 | 205 | Maximum coincidence count is 3 with a hash coincidence efficiency of 73,68421052 63158%. The hash function is classified as BAD / INAPPROPRIATE. | 7 /P=1,035503% | 1 /P=0,147929% | 1 /P=0,147929% | -1 /P=-0,147929% |
| ☑ | ns | 2 | 225 | | 8 /P=1,183432% | 1 /P=0,147929% | 1 /P=0,147929% | -1 /P=-0,147929% |
| ☑ | ts | 2 | 193 | | 18 /P=2,662722% | 1 /P=0,147929% | 1 /P=0,147929% | -1 /P=-0,147929% |
| ☐ | turnleft | 8 | 35 | | | | | |
| ☐ | turnright | 9 | 243 | | | | | |
| ☑ | max | 3 | 167 | | 121 /P=0,6884388% | 11 /P=6,258535E-02% | 7 /P=3,982704E-02% | 0 /P=0% |
| ☑ | min | 3 | 191 | | 171 /P=0,9729176% | 13 /P=0,0739645% | 7 /P=3,982704E-02% | 0 /P=0% |
| ☑ | tcnt | 4 | 99 | | 3607 /P=0,7893193% | 115 /P=2,516543E-02% | 263 /P=5,755226E-02% | 26 /P=5,689577E-03% |
| ☐ | interval | 8 | 207 | ☐ Show results | | | | |

## evaluation of the iterative hash function: H = M Xor ((H + 7 Xor X * 2) / 2)

evaluate hash results — find random strings with the same length

| ☒ | string set | length | hashed | | Text that hashes to the same value | Text that hashes to the same value and has the same starting character | Text that hashes to the same value and has the same ending character | Text that hashes to the same value and has the same 2 ending characters |
|---|---|---|---|---|---|---|---|---|
| | | | | number of coincidences | number of hits / possibility of occurance % (P): | | | |
| ☑ | info | 4 | 183 | **8** | 10895 /P=2,384151% | 456 /P=9,978642E-02% | 1585 /P=0,3468454% | 153 /P=3,348097E-02% |
| ☑ | start | 5 | 172 | | 323451 /P=2,722336% | 12547 /P=0,1056022% | 20243 /P=0,1703759% | 3164 /P=2,662991E-02% |
| ☑ | stop | 4 | 166 | coincidence table: #132:2 #134:2 #140:2 #168:2 | 7440 /P=1,628094% | 310 /P=6,783726E-02% | 1728 /P=0,378138% | 100 /P=2,188299E-02% |
| ☑ | reset | 5 | 168 | | 294380 /P=2,477659% | 10936 /P=9,204321E-02% | 34472 /P=0,2901348% | 3013 /P=2,535902E-02% |
| ☑ | help | 4 | 147 | | 5343 /P=1,169208% | 192 /P=4,201534E-02% | 557 /P=0,1218882% | 21 /P=4,595427E-03% |
| ☐ | destspeed | 9 | 134 | | | | | |
| ☐ | setspeed | 8 | 132 | | | | | |
| ☐ | newspeed | 8 | 132 | | | | | |
| ☐ | targetspeed | 11 | 134 | | | | | |
| ☑ | ds | 2 | 135 | efficiency and classification: | 11 /P=1,627219% | 0 /P=0% | 2 /P=0,295858% | -1 /P=-0,147929% |
| ☑ | ss | 2 | 140 | Maximum coincidence count is 2 with a hash coincidence efficiency of 57,89473684 21053%. The hash function is classified as BAD / INAPPROPRIATE. | 21 /P=3,106509% | 0 /P=0% | 0 /P=0% | -1 /P=-0,147929% |
| ☑ | ns | 2 | 139 | | 15 /P=2,218935% | 0 /P=0% | 2 /P=0,295858% | -1 /P=-0,147929% |
| ☑ | ts | 2 | 143 | | 19 /P=2,810651% | 0 /P=0% | 2 /P=0,295858% | -1 /P=-0,147929% |
| ☐ | turnleft | 8 | 149 | | | | | |
| ☐ | turnright | 9 | 169 | | | | | |
| ☑ | max | 3 | 168 | | 536 /P=3,049613% | 21 /P=0,1194811% | 61 /P=0,3470642% | 4 /P=2,275831E-02% |
| ☑ | min | 3 | 162 | | 606 /P=3,447883% | 25 /P=0,1422394% | 68 /P=0,3868912% | 4 /P=2,275831E-02% |
| ☑ | tcnt | 4 | 150 | | 5568 /P=1,218445% | 192 /P=4,201534E-02% | 1729 /P=0,3783568% | 145 /P=3,173033E-02% |
| ☐ | interval | 8 | 140 | ☐ Show results | | | | |

**A method for command identification, using modified collision free hashing with addition & rotation iterative hash functions. (Part 1)**

### evaluation of the iterative hash function: $H = (M / 2)$ Xor $(H * 2)$

evaluate hash results — find random strings with the same length

| | string set | length | hashed | Count coincidences | Text that hashes to the same value | Text that hashes to the same value and has the same starting character | Text that hashes to the same value and has the same ending character | Text that hashes to the same value and has the same 2 ending characters |
|---|---|---|---|---|---|---|---|---|
| ☑ | info | 4 | 226 | number of coincidences: **5** | 5572 /P=1,21932% | 414 /P=9,059557E-02% | 170 /P=3,720108E-02% | 12 /P=2,625959E-03% |
| ☑ | start | 5 | 56 | | 62872 /P=0,5291643% | 5124 /P=4,312632E-02% | 1858 /P=1,563792E-02% | 368 /P=3,097284E-03% |
| ☑ | stop | 4 | 176 | coincidence table: #142:2 #254:3 | 4796 /P=1,049508% | 374 /P=8,184237E-02% | 372 /P=8,140472E-02% | 26 /P=5,689577E-03% |
| ☑ | reset | 5 | 118 | | 61931 /P=0,5212443% | 5576 /P=4,693059E-02% | 4757 /P=4,003745E-02% | 355 /P=2,987869E-03% |
| ☑ | help | 4 | 252 | | 5776 /P=1,263961% | 438 /P=9,584749E-02% | 414 /P=9,059557E-02% | 30 /P=6,564896E-03% |
| ☐ | destspeed | 9 | 254 | | | | | |
| ☐ | setspeed | 8 | 254 | | | | | |
| ☐ | newspeed | 8 | 190 | | | | | |
| ☐ | targetspeed | 11 | 254 | | | | | |
| ☑ | ds | 2 | 158 | efficiency and classification: Maximum coincidence count is 3 with a hash coincidence efficiency of 73,68421052 63158%. The hash function is classified as BAD / INAPPROPRIATE. | 36 /P=5,325444% | 2 /P=0,295858% | 2 /P=0,295858% | -1 /P=-0,147929% |
| ☑ | ss | 2 | 142 | | 26 /P=3,846154% | 2 /P=0,295858% | 2 /P=0,295858% | -1 /P=-0,147929% |
| ☑ | ns | 2 | 148 | | 32 /P=4,733728% | 2 /P=0,295858% | 0 /P=0% | -1 /P=-0,147929% |
| ☑ | ts | 2 | 142 | | 26 /P=3,846154% | 2 /P=0,295858% | 2 /P=0,295858% | -1 /P=-0,147929% |
| ☐ | turnleft | 8 | 52 | | | | | |
| ☐ | turnright | 9 | 210 | | | | | |
| ☑ | max | 3 | 68 | | 450 /P=2,560309% | 32 /P=0,1820665% | 32 /P=0,1820665% | 2 /P=1,137915E-02% |
| ☑ | min | 3 | 71 | | 150 /P=0,8534365% | 12 /P=6,827492E-02% | 32 /P=0,1820665% | 2 /P=1,137915E-02% |
| ☑ | tcnt | 4 | 140 | | 5003 /P=1,094806% | 438 /P=9,584749E-02% | 125 /P=2,735373E-02% | 24 /P=5,251917E-03% |
| ☐ | interval | 8 | 18 | ☐ Show results | | | | |

### evaluation of the iterative hash function: $H = (M * 2)$ Xor $(H / 2)$

evaluate hash results — find random strings with the same length

| | string set | length | hashed | Count coincidences | Text that hashes to the same value | Text that hashes to the same value and has the same starting character | Text that hashes to the same value and has the same ending character | Text that hashes to the same value and has the same 2 ending characters |
|---|---|---|---|---|---|---|---|---|
| ☑ | info | 4 | 149 | number of coincidences: **10** | 6260 /P=1,369875% | 319 /P=6,980673E-02% | 457 /P=0,1000053% | 35 /P=7,659046E-03% |
| ☑ | start | 5 | 185 | | 91780 /P=0,7724695% | 2820 /P=2,373463E-02% | 7955 /P=6,695352E-02% | 801 /P=6,741643E-03% |
| ☑ | stop | 4 | 168 | coincidence table: #136:3 #149:2 #164:3 #168:2 | 8157 /P=1,784995% | 250 /P=5,470747E-02% | 790 /P=0,1728756% | 45 /P=9,847344E-03% |
| ☑ | reset | 5 | 164 | | 233660 /P=1,966607% | 8167 /P=6,873783E-02% | 22714 /P=0,1911731% | 1240 /P=0,0104365% |
| ☑ | help | 4 | 164 | | 7958 /P=1,741448% | 367 /P=8,031056E-02% | 804 /P=0,1759392% | 57 /P=0,0124733% |
| ☐ | destspeed | 9 | 152 | | | | | |
| ☐ | setspeed | 8 | 136 | | | | | |
| ☐ | newspeed | 8 | 136 | | | | | |
| ☐ | targetspeed | 11 | 139 | | | | | |
| ☑ | ds | 2 | 130 | efficiency and classification: Maximum coincidence count is 3 with a hash coincidence efficiency of 47,36842105 26316%. The hash function is classified as BAD / INAPPROPRIATE. | 9 /P=1,331361% | 0 /P=0% | 0 /P=0% | -1 /P=-0,147929% |
| ☑ | ss | 2 | 149 | | 7 /P=1,035503% | 0 /P=0% | 0 /P=0% | -1 /P=-0,147929% |
| ☑ | ns | 2 | 136 | | 8 /P=1,183432% | 0 /P=0% | 0 /P=0% | -1 /P=-0,147929% |
| ☑ | ts | 2 | 146 | | 7 /P=1,035503% | 0 /P=0% | 0 /P=0% | -1 /P=-0,147929% |
| ☐ | turnleft | 8 | 175 | | | | | |
| ☐ | turnright | 9 | 164 | | | | | |
| ☑ | max | 3 | 168 | | 326 /P=1,854802% | 16 /P=9,103323E-02% | 35 /P=0,1991352% | 2 /P=1,137915E-02% |
| ☑ | min | 3 | 188 | | 374 /P=2,127902% | 18 /P=0,1024124% | 12 /P=6,827492E-02% | 0 /P=0% |
| ☑ | tcnt | 4 | 170 | | 8157 /P=1,784995% | 373 /P=8,162354E-02% | 796 /P=0,1741886% | 60 /P=1,312979E-02% |
| ☐ | interval | 8 | 150 | ☐ Show results | | | | |

**A method for command identification, using modified collision free hashing with addition & rotation iterative hash functions. (Part 1)**

### evaluation of the iterative hash function: H = M Xor (H + 85) Xor (M * 2)

evaluate hash results — find random strings with the same length

| | string set | length | hashed | | Text that hashes to the same value | Text that hashes to the same value and has the same starting character | Text that hashes to the same value and has the same ending character | Text that hashes to the same value and has the same 2 ending characters |
|---|---|---|---|---|---|---|---|---|
| | | | | **Count coincidences** / number of coincidences: **3** | number of hits / possibility of occurance % (P): | | | |
| ☑ | info | 4 | 240 | | 415 /P=0,0908144% | 39 /P=8,534365E-03% | 41 /P=8,972025E-03% | 4 /P=8,753195E-04% |
| ☑ | start | 5 | 133 | | 71987 /P=0,605881% | 2876 /P=2,420595E-02% | 2962 /P=2,492977E-02% | 95 /P=7,995707E-04% |
| ☑ | stop | 4 | 84 | coincidence table: #219:3 | 2938 /P=0,6429222% | 124 /P=0,0271349% | 104 /P=2,275831E-02% | 5 /P=1,094149E-03% |
| ☑ | reset | 5 | 138 | | 72187 /P=0,6075643% | 2941 /P=2,475302E-02% | 2975 /P=2,503919E-02% | 111 /P=9,342352E-04% |
| ☑ | help | 4 | 219 | | 463 /P=0,1013182% | 56 /P=1,225447E-02% | 50 /P=1,094149E-02% | 3 /P=6,564896E-04% |
| ☐ | destspeed | 9 | 42 | | | | | |
| ☐ | setspeed | 8 | 219 | | | | | |
| ☐ | newspeed | 8 | 129 | | | | | |
| ☐ | targetspeed | 11 | 235 | | | | | |
| ☑ | ds | 2 | 219 | efficiency and classification: Maximum coincidence count is 3 with a hash coincidence efficiency of 84,2105263157895%. The hash function is classified as BAD / INAPPROPRIATE. | 3 /P=0,443787% | 0 /P=0% | 0 /P=0% | -1 /P=-0,147929% |
| ☑ | ss | 2 | 128 | | 5 /P=0,7396449% | 0 /P=0% | 0 /P=0% | -1 /P=-0,147929% |
| ☑ | ns | 2 | 169 | | 8 /P=1,183432% | 0 /P=0% | 0 /P=0% | -1 /P=-0,147929% |
| ☑ | ts | 2 | 139 | | 9 /P=1,331361% | 0 /P=0% | 0 /P=0% | -1 /P=-0,147929% |
| ☐ | turnleft | 8 | 158 | | | | | |
| ☐ | turnright | 9 | 206 | | | | | |
| ☑ | max | 3 | 225 | | 119 /P=0,6770597% | 6 /P=3,413746E-02% | 4 /P=2,275831E-02% | 0 /P=0% |
| ☑ | min | 3 | 211 | | 119 /P=0,6770597% | 9 /P=5,120619E-02% | 5 /P=2,844788E-02% | 0 /P=0% |
| ☑ | tcnt | 4 | 107 | | 2836 /P=0,6206015% | 102 /P=2,232065E-02% | 122 /P=2,669724E-02% | 5 /P=1,094149E-03% |
| ☐ | interval | 8 | 253 | Show results | | | | |

### evaluation of the iterative hash function: H = M Xor (H + 170) Xor (M / 2)

evaluate hash results — find random strings with the same length

| | string set | length | hashed | | Text that hashes to the same value | Text that hashes to the same value and has the same starting character | Text that hashes to the same value and has the same ending character | Text that hashes to the same value and has the same 2 ending characters |
|---|---|---|---|---|---|---|---|---|
| | | | | **Count coincidences** / number of coincidences: **0** | number of hits / possibility of occurance % (P): | | | |
| ☑ | info | 4 | 246 | | 919 /P=0,2011047% | 99 /P=2,166416E-02% | 115 /P=2,516543E-02% | 11 /P=2,407129E-03% |
| ☑ | start | 5 | 217 | | 147873 /P=1,244578% | 6593 /P=5,549021E-02% | 5545 /P=4,666968E-02% | 189 /P=1,590725E-03% |
| ☑ | stop | 4 | 172 | coincidence table: | 6708 /P=1,467911% | 300 /P=6,564897E-02% | 181 /P=3,960821E-02% | 9 /P=1,969469E-03% |
| ☑ | reset | 5 | 214 | | 135109 /P=1,137149% | 5633 /P=4,741034E-02% | 6839 /P=5,756067E-02% | 321 /P=2,701707E-03% |
| ☑ | help | 4 | 237 | | 835 /P=0,1827229% | 98 /P=2,144533E-02% | 104 /P=2,275831E-02% | 5 /P=1,094149E-03% |
| ☐ | destspeed | 9 | 183 | | | | | |
| ☐ | setspeed | 8 | 79 | | | | | |
| ☐ | newspeed | 8 | 74 | | | | | |
| ☐ | targetspeed | 11 | 19 | | | | | |
| ☑ | ds | 2 | 111 | efficiency and classification: Maximum coincidence count is 1 with a hash coincidence efficiency of 100%. The hash function is classified as EXCELLENT. | 6 /P=0,887574% | 1 /P=0,147929% | 0 /P=0% | -1 /P=-0,147929% |
| ☑ | ss | 2 | 68 | | 17 /P=2,514793% | 1 /P=0,147929% | 1 /P=0,147929% | -1 /P=-0,147929% |
| ☑ | ns | 2 | 84 | | 13 /P=1,923077% | 1 /P=0,147929% | 0 /P=0% | -1 /P=-0,147929% |
| ☑ | ts | 2 | 71 | | 10 /P=1,47929% | 1 /P=0,147929% | 0 /P=0% | -1 /P=-0,147929% |
| ☐ | turnleft | 8 | 1 | | | | | |
| ☐ | turnright | 9 | 98 | | | | | |
| ☑ | max | 3 | 112 | | 188 /P=1,06964% | 9 /P=5,120619E-02% | 13 /P=0,0739645% | 0 /P=0% |
| ☑ | min | 3 | 105 | | 243 /P=1,382567% | 22 /P=0,1251707% | 18 /P=0,1024124% | 0 /P=0% |
| ☑ | tcnt | 4 | 180 | | 6572 /P=1,43815% | 204 /P=4,464129E-02% | 204 /P=4,464129E-02% | 8 /P=1,750639E-03% |
| ☐ | interval | 8 | 80 | Show results | | | | |

**evaluation of the iterative hash function: H = M Xor (M + 170) Xor (H / 2)**

| | string set | length | hashed | number of coincidences | Text that hashes to the same value | Text that hashes to the same value and has the same starting character | Text that hashes to the same value and has the same ending character | Text that hashes to the same value and has the same 2 ending characters |
|---|---|---|---|---|---|---|---|---|
| ☑ | info | 4 | 94 | 12 | 9412 /P=2,059627% | 395 /P=0,0864378% | 697 /P=0,1525244% | 61 /P=1,334862E-02% |
| ☑ | start | 5 | 76 | | 463433 /P=3,900499% | 17763 /P=0,1495029% | 35333 /P=0,2973814% | 2787 /P=2,345688E-02 |
| ☑ | stop | 4 | 70 | coincidence table: | 12293 /P=2,690076% | 545 /P=0,1192623% | 1017 /P=0,22255% | 71 /P=1,553692E-02% |
| ☑ | reset | 5 | 79 | #65:3 #70:2 #76:5 #91:2 | 151815 /P=1,277756% | 5791 /P=4,874014E-02 | 11839 /P=9,964334E-02 | 857 /P=7,212969E-03% |
| ☑ | help | 4 | 65 | | 11125 /P=2,434482% | 403 /P=8,818844E-02% | 705 /P=0,1542751% | 55 /P=1,203564E-02% |
| ☐ | destspeed | 9 | 76 | | | | | |
| ☐ | setspeed | 8 | 76 | | | | | |
| ☐ | newspeed | 8 | 76 | | | | | |
| ☐ | targetspeed | 11 | 76 | | | | | |
| ☑ | ds | 2 | 91 | efficiency and classification: | 35 /P=5,177515% | 3 /P=0,443787% | 6 /P=0,887574% | -1 /P=-0,147929% |
| ☑ | ss | 2 | 89 | Maximum coincidence count is 5 with a hash coincidence efficiency of 36,84210526 31579%. The hash function is classified as BAD / INAPPROPRIATE. | 19 /P=2,810651% | 3 /P=0,443787% | 3 /P=0,443787% | -1 /P=-0,147929% |
| ☑ | ns | 2 | 85 | | 37 /P=5,473373% | 3 /P=0,443787% | 3 /P=0,443787% | -1 /P=-0,147929% |
| ☑ | ts | 2 | 91 | | 35 /P=5,177515% | 3 /P=0,443787% | 6 /P=0,887574% | -1 /P=-0,147929% |
| ☐ | turnleft | 8 | 65 | | | | | |
| ☐ | turnright | 9 | 65 | | | | | |
| ☑ | max | 3 | 118 | | 293 /P=1,667046% | 13 /P=0,0739645% | 55 /P=0,3129267% | 3 /P=1,706873E-02% |
| ☑ | min | 3 | 82 | | 748 /P=4,255804% | 29 /P=0,1649977% | 45 /P=0,2560309% | 3 /P=1,706873E-02% |
| ☑ | tcnt | 4 | 70 | | 12293 /P=2,690076% | 537 /P=0,1175116% | 1017 /P=0,22255% | 71 /P=1,553692E-02% |
| ☐ | interval | 8 | 80 | ☐ Show results | | | | |

Buttons: hash the string set | evaluate hash results | Count coincidences | find random strings with the same length

**Comments**

The hash size is very small (1 byte), so many conflicts (collisions) occur. The first criterion (of same length) is therefore necessary. The results are presented in the first column.

It is assumed that the possibility that a string of certain length will occur at the input is equal to the possibility that a string of any other length will occur. If this assumption is not correct for an application, the efficiency calculations must have weighting factors multiplied with each possibility for each length.

The classification of the hash functions presented at the first column regards only intrinsic collisions between valid strings of any length. If the same-length condition is to be met, this classification is of no importance unless there is an intrinsic conflict between two valid strings of the *same* length.

After this evaluation analysis, one could identify three basic addition & rotation hash function categories:
i) The non-destructive category: these functions operate on the blocks without applying changes to many bits – therefore many strings of the same length hash to same values. An example is the iterative hash function $H=H \oplus M$, which allows many permutations to hash to the same value – the conflict possibility is therefore (as shown in the above figures) one of the highest. Shift instructions are destructive, so they do not appear in hash functions of this category.
ii) The destructive category: The final result depends mostly on the first or the last blocks ( as an example, $H=H \oplus (H\&M)$ or $H=M \oplus (M+170) \oplus (H>>1)$).
iii) The intermediate category: these functions have an intermediate effect on the blocks, having both non-destructive and destructive characteristics. They achieve the best performance and are therefore recommended. Example: $H=M \oplus (H+85) \oplus (M<<1)$

The evaluation will not be complete unless one calculates the overall factor K of extrinsic conflict and also the maximum extrinsic collision possibility for a certain string length: max(P | length).
K is given by the following formula:

$$K = \sum_{length} P(length)$$

The overall factor K is the sum the possibilities for each string length (including all strings). For example, in our string set evaluated above

K = P(2) + P(3) + P(4) + P(5), where

P(2) = P(«ds»)+P(«ss»)+P(«ns»)+P(«ts»),

P(3) = P(«max»)+P(«min»), etc.

Possibilities concerning strings of greater length need too much time to be calculated (the time consumed by the algorithm changes exponentially with a change of *length*) and also have very small values, so they can be ommited.

max(P | length) is given by the formula:

$$\max(P \mid length) = P(length),$$

that is the maximum collision possibility for a certain length. Example: for length=2, max(P | 2) = P(2) = P(«ds»)+P(«ss»)+P(«ns»)+P(«ts»)

Factor K is used to compare the different hash functions. The value of max(P) gives us an idea of how well the function will behave.

To conclude, the best iterative addition & rotation hash function for the given string set, achieving an overall conflict factor K equal to 7.7213%, is the function H=M⊕(H+85) ⊕(M<<1).

### Discussion

The hash functions presented here are targeted towards a kind of table-lookup*. They have a very small block size and therefore are not intended for use in any kind of cryptographic/security applications. Cryptographic hash functions require a random and uniform mapping of input values to output/hashed values and also require very strict and thorough evaluation analysis.

On the other hand, the desired operation of the additive & rotating hash functions in this paper is a non-uniform distribution of results; non-uniform in a way that the valid command strings are mapped into their own values, while all other invalid command strings are mapped on completely different values (no or very small collision with the hashed valid strings). But is such a distribution possible to achieve? Such a distribution exists (the interested reader could refer to the common digital design techniques, i.e. the Quine – McCluskey method for logic minimization). However, the implementation of such a solution would probably be too slow and/or require much code-size, rendering its usage in general purpose microcontrollers impractical. It should be noted that such a distribution is not achieved with addition & rotation hash functions.

Nevertheless, the mapping of the hashed invalid strings does not need to be random; only a low-possibility of extrinsic collision is important. It is also required that the hashed valid strings are not the same. (no intrinsic collision). An example of this is shown in the following figure. Strings in bold indicate valid commands.

---

\* but in any way not for a large dictionary table, mainly because of the very small hash length. If hashing was ideal, at most $256^z$ valid strings/commands would be mapped into their unique hash values, where z is the length (in bytes) of the hash result.

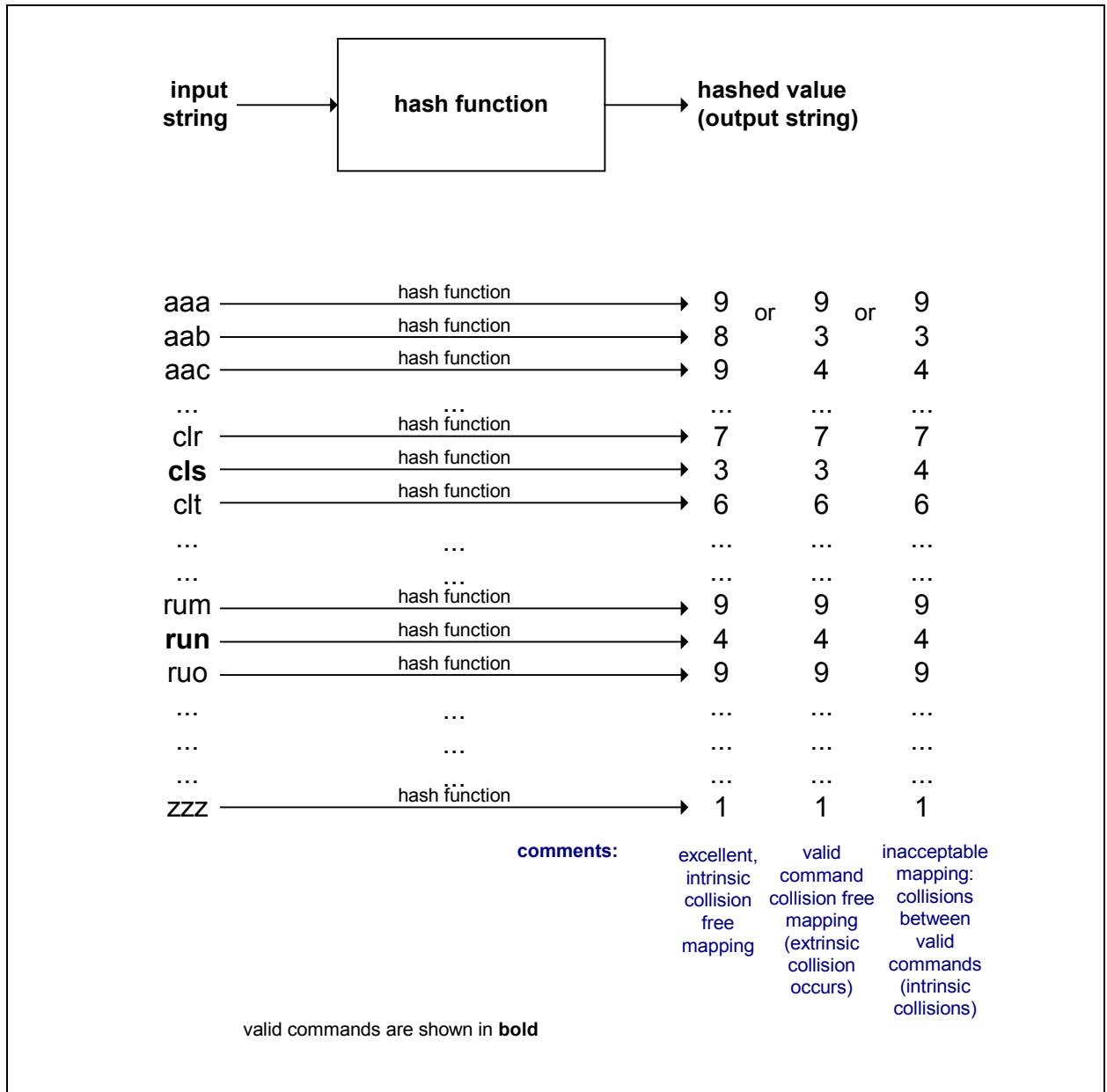

**Figure 1.** Example of three different hash mappings.

## Appendix

The following program is a sample visual basic program that can be used to evaluate the iterative addition & rotation hash functions for string set identification. It was written in visual basic 5 and the project files are found inside the zip file accompanying this paper.

CODE for HASHEVAL.FRM:

```
'______________________________________________
'
' Constants:
'______________________________________________
Const detectcollision = False             ' Enable or disable to detect or not the special
collision described in the paper

Const chr_start = 225                     ' the starting character and
Const chr_stop = 250                      ' the stoping character used to create the random
strings

Const Hstart = 0                          ' initial feed for the hash function
```

```
Const coincidence_limit = 5                         ' % : used to characterize results after
calculating coincidence of the hash results
Const coinc_mes_BAD = "BAD / INAPPROPRIATE"
Const coinc_mes_NOTBAD = "NOT GENERALLY RECOMMENDED"    'if smaller than the limit
Const coinc_mes_GOOD = "EXCELLENT"

Const offset = 128                          ' converts between ascii and lcd character

'________________________________________________
'
' Globals:
'________________________________________________

Dim hashname                                'the hash function used (this is a string that is
shown on the screen)
Dim i                                       'general purpose

Dim showresults                             'true/false : print results of random string
hashes or not

'used in the  hashing of random strings  procedure
Dim text2hash
Dim length
Dim same_hash_counter
Dim buffer_chr          'used for storing a character which is desired to be present in the
random string
Dim buffer_chr2         'used for storing another character which is desired to be present in
the random string

Dim occurence_p As Single 'stores the hash result occurence possibility between a random
string and the original string
Dim secondcol           'counter for the special collision described in the paper

' "M" is the input, "H" is the output of the hash function:
Dim H
Dim M

' "result" stores the results of the hashing of the text in the text boxes.
' The dimension is configured later
Dim result

Dim coincidence_table(255)

' A switch to make the user's life easier when selecting instructions
' function: select all / none instructions
Dim select_instructions_switch

' The number of texts to hash
Dim instruction_number
Private Sub calculate_hash_value()
    'H = M Xor (H And M)
    'hashname = "H = M Xor (H And M)"
    'H = M Xor M / 2 Xor (H + 1)
    'hashname = "H = M Xor M / 2 Xor (H + 1)"
    'H = M Xor (M / 4) Xor (H + 1)
    'hashname = "H = M Xor (M / 4) Xor (H + 1)"
    'H = M Xor (M / 2) Xor (H + 2)
    'hashname = "H = M Xor (M / 2) Xor (H + 2)"
    'H = M Xor (M / 2) Xor (H * 2)
    'hashname = "H = M Xor (M / 2) Xor (H * 2)"
    'H = M Xor H
    'hashname = "H = M Xor H"

    'H = M Xor ((H + 7 Xor X * 2) / 2)
    'hashname = "H = M Xor ((H + 7 Xor X * 2) / 2)"

    'H = M Xor (H + 170) Xor (M / 2)
    'hashname = "H = M Xor (H + 170) Xor (M / 2)"
    'H = M Xor (H + 85) Xor (M * 2)
    'hashname = "H = M Xor (H + 85) Xor (M * 2)"
    'H = M Xor (M + 170) Xor (H / 2)          ' (similar hash values for similar commands)
    'hashname = "H = M Xor (M + 170) Xor (H / 2)"
    'H = M Xor (M + 85) Xor (H * 2)
    'hashname = "H = M Xor (M + 85) Xor (H * 2)"

    'H = (M / 2) Xor (H * 2)
    'hashname = "H = (M / 2) Xor (H * 2)"

    H = (M * 2) Xor (H / 2)
    hashname = "H = (M * 2) Xor (H / 2)"
```

```
            'If overflow, bits higher than the 7th are lost
            If H > 255 Then
                'MsgBox ("before: " & H)
                While (H > 0)
                    H = H - 256
                Wend
                If (H < 0) Then H = H + 256
                'MsgBox ("after: " & H)
            End If
End Sub
Private Sub show_hash_results()
    For i = 0 To instruction_number                   'hash all instructions
        Label2(i).Caption = result(i)
    Next i
End Sub
Private Sub calculate_hash_results()
    For i = 0 To instruction_number                   'hash all instructions
        H = Hstart                                    'Initial hash value is defined in the
declarations section

        For j = 1 To Len(Text1(i).Text)               'read every character from the text
            M = Asc(Mid$(Text1(i).Text, j, 1)) + offset    'read input byte. +offset converts
ASCII to LCD character set
            calculate_hash_value
        Next j

    result(i) = H                                     'store hash results
    Next i
End Sub

Private Sub Check2_Click()
If showresults = False Then showresults = True Else showresults = False
End Sub

Private Sub visualize()
'count the length of all strings
    For abc = 0 To instruction_number
    Text9(abc).Text = Len(Text1(abc).Text)
    Text9(abc).ForeColor = QBColor(Len(Text1(abc).Text))
    'Text11(abc).BackColor = QBColor(Val(Text11(abc).Text) Mod 16)
    Next abc

End Sub
Private Sub Command1_Click()
    'count the length of all strings in case the user changed the strings
    visualize
    calculate_hash_results
    show_hash_results
End Sub

Private Sub initialize_values()
    ' The initial value for selecting/deselecting text boxes
    select_instructions_switch = 0
    ' The number of texts to hash
    instruction_number = Text1.Count - 1

    ' Reserve hash function results table
    ReDim result(instruction_number)

    ' do not show random-string hash matches
    showresults = 0

    secondcol = 0

    visualize
    End Sub

Private Sub count_hash_results_coincidences()
    'clear the coincidence table
    For i = 0 To 255
        coincidence_table(i) = 0
    Next i

    'record coincidences
    For i = 0 To instruction_number
        coincidence_table(result(i)) = coincidence_table(result(i)) + 1
    Next i

    'find maximum counts
    Max = 0
    For i = 0 To 255
```

```
            If coincidence_table(i) > Max Then Max = coincidence_table(i)
    Next i

    'Evaluate hash results
    count_sum = 0
    For i = 0 To 255
        If coincidence_table(i) > 1 Then
        count_sum = count_sum + coincidence_table(i)
        End If
    Next i

    hash_coincidence_ratio = count_sum / (instruction_number + 1) '+1 because it counts from 0
    hash_coincidence_ratio = hash_coincidence_ratio * 100         ' calculate % value

    If hash_coincidence_ratio >= coincidence_limit Then coinc_mes = coinc_mes_BAD
    If hash_coincidence_ratio < coincidence_limit Then coinc_mes = coinc_mes_NOTBAD
    If hash_coincidence_ratio = 0 Then coinc_mes = coinc_mes_GOOD

    'print results
    result_string = ""
    For i = 0 To 255
        If coincidence_table(i) > 1 Then 'show only if there is a coincidence
        result_string = result_string & " #" & i & ":" & coincidence_table(i)
        End If
    Next i
    Text6.Text = result_string

    Text7.Text = "Maximum coincidence count is " & Max & " with a hash coincidence efficiency
of " & (100 - hash_coincidence_ratio) & "%" & ". The hash function is classified as " &
coinc_mes & "."

    Text2.Text = count_sum
End Sub

Private Sub Command2_Click()
    count_hash_results_coincidences
End Sub

Private Sub hash_random_string()
' Sorry for the dumm code here but this is the faster (and easier) solution
' DETECTION FOR 2nd TYPE COLLISION IS ENABLED ONLY FOR INSTRUCTIONS OF LENGTH 4 AND LESS
' (due to practical reasons)

If length = 1 Then
            For j0 = chr_start To chr_stop

                H = Hstart
                text2hash(0) = j0

                For j = 0 To length - 1
                    M = text2hash(j)
                    calculate_hash_value
                Next j
                                'MsgBox (text2hash(0) & " " & text2hash(1) & " " &
text2hash(2) & " " & text2hash(3) & " " & text2hash(4) & " " & " HASH VALUE=" & H)
        If H = result(i) Then
                                same_hash_counter = same_hash_counter + 1
                                If showresults = True Then
                                If detectcollision Then
                                    For abc = 0 To instruction_number
                                        If Left$(Text1(abc).Text, length) = Chr$(text2hash(0) -
offset) Then
                                                        MsgBox ("OOPS. 2nd type collision with
" & Text1(abc).Text)
                                                        secondcol = secondcol + 1
                                        End If
                                    Next abc
                                End If
                                MsgBox (Chr$(text2hash(0) - offset) & " HASH VALUE=" & H)
                                End If
                    End If
            Next
End If

If length = 2 Then
            For j0 = chr_start To chr_stop
            For j1 = chr_start To chr_stop

                H = Hstart
                text2hash(0) = j0
                text2hash(1) = j1

                For j = 0 To length - 1
                    M = text2hash(j)
```

```
                                calculate_hash_value
                        Next j
                                        'MsgBox (text2hash(0) & " " & text2hash(1) & " " &
text2hash(2) & " " & text2hash(3) & " " & text2hash(4) & " " & " HASH VALUE=" & H)
        If H = result(i) Then
                                        same_hash_counter = same_hash_counter + 1

                                If detectcollision Then
                                    For abc = 0 To instruction_number
                                        If Left$(Text1(abc).Text, length) = (Chr$(text2hash(0) -
offset) & Chr$(text2hash(1) - offset)) Then
                                            MsgBox ("OOPS. 2nd type collision
with " & Text1(abc).Text)
                                            secondcol = secondcol + 1
                                        End If

                                    Next abc
                                End If

                                If showresults = True Then
                                    MsgBox (Chr$(text2hash(0) - offset) & " " & Chr$(text2hash(1)
- offset) & " HASH VALUE=" & H)
                                End If
                        End If
            Next
            Next
End If

If length = 3 Then
            For j0 = chr_start To chr_stop
            For j1 = chr_start To chr_stop
            For j2 = chr_start To chr_stop

                H = Hstart
                text2hash(0) = j0
                text2hash(1) = j1
                text2hash(2) = j2

                For j = 0 To length - 1
                    M = text2hash(j)
                    calculate_hash_value
                Next j
                                        'MsgBox (text2hash(0) & " " & text2hash(1) & " " &
text2hash(2) & " " & text2hash(3) & " " & text2hash(4) & " " & " HASH VALUE=" & H)
        If H = result(i) Then
                                        same_hash_counter = same_hash_counter + 1

                                If detectcollision Then
                                    For abc = 0 To instruction_number
                                        If Left$(Text1(abc).Text, length) = (Chr$(text2hash(0) -
offset) & Chr$(text2hash(1) - offset) & Chr$(text2hash(2) - offset)) Then
                                            MsgBox ("OOPS. 2nd type collision with
" & Text1(abc).Text)
                                            secondcol = secondcol + 1
                                        End If
                                    Next abc
                                End If

                                If showresults = True Then
                                    MsgBox (Chr$(text2hash(0) - offset) & " " & Chr$(text2hash(1)
- offset) & " " & Chr$(text2hash(2) - offset) & " " & " HASH VALUE=" & H)
                                End If
                        End If

            Next
            Next
            Next
End If

If length = 4 Then
            For j0 = chr_start To chr_stop
            For j1 = chr_start To chr_stop
            For j2 = chr_start To chr_stop
            For j3 = chr_start To chr_stop

                H = Hstart
                text2hash(0) = j0
                text2hash(1) = j1
                text2hash(2) = j2
                text2hash(3) = j3

                For j = 0 To length - 1
                    M = text2hash(j)
```

```
                            calculate_hash_value
                    Next j
                                    'MsgBox  (text2hash(0)  &  "  "  &  text2hash(1)  &  "  "  &
text2hash(2) & " " & text2hash(3) & " " & text2hash(4) & " " & " HASH VALUE=" & H)
        If H = result(i) Then
                                    same_hash_counter = same_hash_counter + 1

                                If detectcollision Then
                                    For abc = 0 To instruction_number
                                        If Left$(Text1(abc).Text, length) = (Chr$(text2hash(0) -
offset) & Chr$(text2hash(1) - offset) & Chr$(text2hash(2) - offset) & Chr$(text2hash(3) -
offset)) Then
                                            MsgBox ("OOPS. 2nd type collision with " &
Text1(abc).Text)
                                            secondcol = secondcol + 1
                                        End If

                                    Next abc
                                End If

                                If showresults = True Then
                                    MsgBox (Chr$(text2hash(0) - offset) & " " & Chr$(text2hash(1)
- offset) & " " & Chr$(text2hash(2) - offset) & " " & Chr$(text2hash(3) - offset) & " HASH
VALUE=" & H)
                                End If
                    End If
            Next
            Next
            Next
            Next
End If

If length = 5 Then
            For j0 = chr_start To chr_stop
            For j1 = chr_start To chr_stop
            For j2 = chr_start To chr_stop
            For j3 = chr_start To chr_stop
            For j4 = chr_start To chr_stop

                H = Hstart
                text2hash(0) = j0
                text2hash(1) = j1
                text2hash(2) = j2
                text2hash(3) = j3
                text2hash(4) = j4
                For j = 0 To length - 1
                    M = text2hash(j)
                    calculate_hash_value
                Next j
                                    'MsgBox  (text2hash(0)  &  "  "  &  text2hash(1)  &  "  "  &
text2hash(2) & " " & text2hash(3) & " " & text2hash(4) & " " & " HASH VALUE=" & H)
        If H = result(i) Then
                                    same_hash_counter = same_hash_counter + 1
                                    If showresults = True Then MsgBox (Chr$(text2hash(0) - offset)
&  "  "  &  Chr$(text2hash(1)  -  offset)  &  "  "  &  Chr$(text2hash(2)  -  offset)  &  "  "  &
Chr$(text2hash(3) - offset) & " " & Chr$(text2hash(4) - offset) & " HASH VALUE=" & H)
                    End If
            Next
            Next
            Next
            Next
            Next
End If

If length = 6 Then
            For j0 = chr_start To chr_stop
            For j1 = chr_start To chr_stop
            For j2 = chr_start To chr_stop
            For j3 = chr_start To chr_stop
            For j4 = chr_start To chr_stop
            For j5 = chr_start To chr_stop

                H = Hstart
                text2hash(0) = j0
                text2hash(1) = j1
                text2hash(2) = j2
                text2hash(3) = j3
                text2hash(4) = j4
                text2hash(5) = j5
                For j = 0 To length - 1
                    M = text2hash(j)
                    calculate_hash_value
                Next j
```

```
                                    'MsgBox  (text2hash(0)  &  "  "  &  text2hash(1)  &  "  "  &
text2hash(2) & " " & text2hash(3) & " " & text2hash(4) & " " & " HASH VALUE=" & H)
        If H = result(i) Then
                                    same_hash_counter = same_hash_counter + 1
                                    If showresults = True Then MsgBox (Chr$(text2hash(0) - offset)
& " " & Chr$(text2hash(1) - offset) & " " & Chr$(text2hash(2) - offset) & " " &
Chr$(text2hash(3) - offset) & " " & Chr$(text2hash(4) - offset) & " " & Chr$(text2hash(5) -
offset) & " HASH VALUE=" & H)
                                End If
            Next
            Next
            Next
            Next
            Next
            Next
End If

If length = 7 Then
            For j0 = chr_start To chr_stop
            For j1 = chr_start To chr_stop
            For j2 = chr_start To chr_stop
            For j3 = chr_start To chr_stop
            For j4 = chr_start To chr_stop
            For j5 = chr_start To chr_stop
            For j6 = chr_start To chr_stop

                H = Hstart
                text2hash(0) = j0
                text2hash(1) = j1
                text2hash(2) = j2
                text2hash(3) = j3
                text2hash(4) = j4
                text2hash(5) = j5
                text2hash(6) = j6
                For j = 0 To length - 1
                    M = text2hash(j)
                    calculate_hash_value
                Next j
                                    'MsgBox  (text2hash(0)  &  "  "  &  text2hash(1)  &  "  "  &
text2hash(2) & " " & text2hash(3) & " " & text2hash(4) & " " & " HASH VALUE=" & H)
        If H = result(i) Then
                                    same_hash_counter = same_hash_counter + 1
                                    If showresults = True Then MsgBox (Chr$(text2hash(0) - offset)
& " " & Chr$(text2hash(1) - offset) & " " & Chr$(text2hash(2) - offset) & " " &
Chr$(text2hash(3) - offset) & " " & Chr$(text2hash(4) - offset) & " " & Chr$(text2hash(5) -
offset) & " " & Chr$(text2hash(6) - offset) & " HASH VALUE=" & H)
                                End If
            Next
            Next
            Next
            Next
            Next
            Next
            Next
End If

If length = 8 Then
            For j0 = chr_start To chr_stop
            For j1 = chr_start To chr_stop
            For j2 = chr_start To chr_stop
            For j3 = chr_start To chr_stop
            For j4 = chr_start To chr_stop
            For j5 = chr_start To chr_stop
            For j6 = chr_start To chr_stop
            For j7 = chr_start To chr_stop

                H = Hstart
                text2hash(0) = j0
                text2hash(1) = j1
                text2hash(2) = j2
                text2hash(3) = j3
                text2hash(4) = j4
                text2hash(5) = j5
                text2hash(6) = j6
                text2hash(7) = j7
                For j = 0 To length - 1
                    M = text2hash(j)
                    calculate_hash_value
                Next j
                                    'MsgBox  (text2hash(0)  &  "  "  &  text2hash(1)  &  "  "  &
text2hash(2) & " " & text2hash(3) & " " & text2hash(4) & " " & " HASH VALUE=" & H)
        If H = result(i) Then
                                    same_hash_counter = same_hash_counter + 1
```

```
                                      If showresults = True Then MsgBox (Chr$(text2hash(0) - offset)
& " " & Chr$(text2hash(1) - offset) & " " & Chr$(text2hash(2) - offset) & " " &
Chr$(text2hash(3) - offset) & " " & Chr$(text2hash(4) - offset) & " " & Chr$(text2hash(5) -
offset) & " " & Chr$(text2hash(6) - offset) & " " & Chr$(text2hash(7) - offset) & " HASH
VALUE=" & H)
                            End If
            Next
            Next
            Next
            Next
            Next
            Next
            Next
            Next
End If

' MAKING CODE FOR LENGTHS OF 9 or MORE IS NOT APPLICABLE
' A length of 9 results in several hours of computation
' 10: several days
' 11: several months
' 12: several years

End Sub
Private Sub Command3_Click()

secondcol = 0

For i = 0 To instruction_number
        If Check1(i).Value = False Then GoTo next_i 'do this only if the box next to the
instruction is checked
    same_hash_counter = 0

        ' Create random strings of the same length and compare the hash results
        length = Len(Text1(i).Text)
        ReDim text2hash(length)

            hash_random_string 'with same length
            same_hash_counter = same_hash_counter - 1 ' because it counted the original string
too
            occurence_p = ((same_hash_counter / (chr_stop - chr_start + 1) ^ length) * 100)
            Text3(i).Text = same_hash_counter & " /P=" & occurence_p & "%"

            special_collision_hits = secondcol - instructionnumber
            If detectcollision Then
            Cls
            Print "Special collision counter="; special_collision_hits
            End If
next_i:
    Next i
End Sub
Private Sub hash_random_string_same2Cstop()
If length = 3 Then
            For j0 = chr_start To chr_stop
            j1 = buffer_chr2
            j2 = buffer_chr

                H = Hstart
                text2hash(0) = j0
                text2hash(1) = j1
                text2hash(2) = j2

                For j = 0 To length - 1
                    M = text2hash(j)
                    calculate_hash_value
                Next j
                                    'MsgBox (text2hash(0) & " " & text2hash(1) & " " &
text2hash(2) & " " & text2hash(3) & " " & text2hash(4) & " " & " HASH VALUE=" & H)
        If H = result(i) Then
                                    same_hash_counter = same_hash_counter + 1
                                    If showresults = True Then MsgBox (Chr$(text2hash(0) - offset)
& " " & Chr$(text2hash(1) - offset) & " " & Chr$(text2hash(2) - offset) & " " & " HASH VALUE="
& H)
                    End If

            Next
End If

If length = 4 Then
            For j0 = chr_start To chr_stop
            For j1 = chr_start To chr_stop
            j2 = buffer_chr2
            j3 = buffer_chr
```

```
                    H = Hstart
                    text2hash(0) = j0
                    text2hash(1) = j1
                    text2hash(2) = j2
                    text2hash(3) = j3

                    For j = 0 To length - 1
                        M = text2hash(j)
                        calculate_hash_value
                    Next j
                                    'MsgBox  (text2hash(0)  &  "  "  &  text2hash(1)  &  "  "  &
text2hash(2) & " " & text2hash(3) & " " & text2hash(4) & " " & " HASH VALUE=" & H)
        If H = result(i) Then
                                    same_hash_counter = same_hash_counter + 1
                                    If showresults = True Then MsgBox (Chr$(text2hash(0) - offset)
& " " & Chr$(text2hash(1) - offset) & " " & Chr$(text2hash(2) - offset) & " " &
Chr$(text2hash(3) - offset) & " HASH VALUE=" & H)
                        End If
            Next
            Next
End If

If length = 5 Then
            For j0 = chr_start To chr_stop
            For j1 = chr_start To chr_stop
            For j2 = chr_start To chr_stop
            j3 = buffer_chr2
            j4 = buffer_chr

                    H = Hstart
                    text2hash(0) = j0
                    text2hash(1) = j1
                    text2hash(2) = j2
                    text2hash(3) = j3
                    text2hash(4) = j4
                    For j = 0 To length - 1
                        M = text2hash(j)
                        calculate_hash_value
                    Next j
                                    'MsgBox  (text2hash(0)  &  "  "  &  text2hash(1)  &  "  "  &
text2hash(2) & " " & text2hash(3) & " " & text2hash(4) & " " & " HASH VALUE=" & H)
        If H = result(i) Then
                                    same_hash_counter = same_hash_counter + 1
                                    If showresults = True Then MsgBox (Chr$(text2hash(0) - offset)
& " " & Chr$(text2hash(1) - offset) & " " & Chr$(text2hash(2) - offset) & " " &
Chr$(text2hash(3) - offset) & " " & Chr$(text2hash(4) - offset) & " HASH VALUE=" & H)
                        End If
            Next
            Next
            Next
End If

If length = 6 Then
            For j0 = chr_start To chr_stop
            For j1 = chr_start To chr_stop
            For j2 = chr_start To chr_stop
            For j3 = chr_start To chr_stop
            j4 = buffer_chr2
            j5 = buffer_chr

                    H = Hstart
                    text2hash(0) = j0
                    text2hash(1) = j1
                    text2hash(2) = j2
                    text2hash(3) = j3
                    text2hash(4) = j4
                    text2hash(5) = j5
                    For j = 0 To length - 1
                        M = text2hash(j)
                        calculate_hash_value
                    Next j
                                    'MsgBox  (text2hash(0)  &  "  "  &  text2hash(1)  &  "  "  &
text2hash(2) & " " & text2hash(3) & " " & text2hash(4) & " " & " HASH VALUE=" & H)
        If H = result(i) Then
                                    same_hash_counter = same_hash_counter + 1
                                    If showresults = True Then MsgBox (Chr$(text2hash(0) - offset)
& " " & Chr$(text2hash(1) - offset) & " " & Chr$(text2hash(2) - offset) & " " &
Chr$(text2hash(3) - offset) & " " & Chr$(text2hash(4) - offset) & " " & Chr$(text2hash(5) -
offset) & " HASH VALUE=" & H)
                        End If
            Next
            Next
            Next
```

```
            Next
End If

If length = 7 Then
        For j0 = chr_start To chr_stop
        For j1 = chr_start To chr_stop
        For j2 = chr_start To chr_stop
        For j3 = chr_start To chr_stop
        For j4 = chr_start To chr_stop
        j5 = buffer_chr2
        j6 = buffer_chr

            H = Hstart
            text2hash(0) = j0
            text2hash(1) = j1
            text2hash(2) = j2
            text2hash(3) = j3
            text2hash(4) = j4
            text2hash(5) = j5
            text2hash(6) = j6
            For j = 0 To length - 1
                M = text2hash(j)
                calculate_hash_value
            Next j
                                'MsgBox (text2hash(0) & " " & text2hash(1) & " " &
text2hash(2) & " " & text2hash(3) & " " & text2hash(4) & " " & " HASH VALUE=" & H)
        If H = result(i) Then
                            same_hash_counter = same_hash_counter + 1
                            If showresults = True Then MsgBox (Chr$(text2hash(0) - offset)
& " " & Chr$(text2hash(1) - offset) & " " & Chr$(text2hash(2) - offset) & " " &
Chr$(text2hash(3) - offset) & " " & Chr$(text2hash(4) - offset) & " " & Chr$(text2hash(5) -
offset) & " " & Chr$(text2hash(6) - offset) & " HASH VALUE=" & H)
                        End If
        Next
        Next
        Next
        Next
        Next
End If

If length = 8 Then
        For j0 = chr_start To chr_stop
        For j1 = chr_start To chr_stop
        For j2 = chr_start To chr_stop
        For j3 = chr_start To chr_stop
        For j4 = chr_start To chr_stop
        For j5 = chr_start To chr_stop
        j6 = buffer_chr2
        j7 = buffer_chr

            H = Hstart
            text2hash(0) = j0
            text2hash(1) = j1
            text2hash(2) = j2
            text2hash(3) = j3
            text2hash(4) = j4
            text2hash(5) = j5
            text2hash(6) = j6
            text2hash(7) = j7
            For j = 0 To length - 1
                M = text2hash(j)
                calculate_hash_value
            Next j
                                'MsgBox (text2hash(0) & " " & text2hash(1) & " " &
text2hash(2) & " " & text2hash(3) & " " & text2hash(4) & " " & " HASH VALUE=" & H)
        If H = result(i) Then
                            same_hash_counter = same_hash_counter + 1
                            If showresults = True Then MsgBox (Chr$(text2hash(0) - offset)
& " " & Chr$(text2hash(1) - offset) & " " & Chr$(text2hash(2) - offset) & " " &
Chr$(text2hash(3) - offset) & " " & Chr$(text2hash(4) - offset) & " " & Chr$(text2hash(5) -
offset) & " " & Chr$(text2hash(6) - offset) & " " & Chr$(text2hash(7) - offset) & " HASH
VALUE=" & H)
                        End If
            Next
            Next
            Next
            Next
            Next
            Next
End If

End Sub
```

```vb
Private Sub hash_random_string_samestop()
If length = 2 Then
        For j0 = chr_start To chr_stop
        j1 = buffer_chr
        
            H = Hstart
            text2hash(0) = j0
            text2hash(1) = j1
            
            For j = 0 To length - 1
                M = text2hash(j)
                calculate_hash_value
            Next j
                                        'MsgBox  (text2hash(0)  &  "  "  &  text2hash(1)  &  "  "  &
text2hash(2) & " " & text2hash(3) & " " & text2hash(4) & " " & " HASH VALUE=" & H)
        If H = result(i) Then
                                        same_hash_counter = same_hash_counter + 1
                                        If showresults = True Then MsgBox (Chr$(text2hash(0) - offset)
& " " & Chr$(text2hash(1) - offset) & " HASH VALUE=" & H)
                        End If
        Next
End If

If length = 3 Then
        For j0 = chr_start To chr_stop
        For j1 = chr_start To chr_stop
        j2 = buffer_chr
        
            H = Hstart
            text2hash(0) = j0
            text2hash(1) = j1
            text2hash(2) = j2
            
            For j = 0 To length - 1
                M = text2hash(j)
                calculate_hash_value
            Next j
                                        'MsgBox  (text2hash(0)  &  "  "  &  text2hash(1)  &  "  "  &
text2hash(2) & " " & text2hash(3) & " " & text2hash(4) & " " & " HASH VALUE=" & H)
        If H = result(i) Then
                                        same_hash_counter = same_hash_counter + 1
                                        If showresults = True Then MsgBox (Chr$(text2hash(0) - offset)
& " " & Chr$(text2hash(1) - offset) & " " & Chr$(text2hash(2) - offset) & " " & " HASH VALUE="
& H)
                        End If
        
        Next
        Next
End If

If length = 4 Then
        For j0 = chr_start To chr_stop
        For j1 = chr_start To chr_stop
        For j2 = chr_start To chr_stop
        j3 = buffer_chr
        
            H = Hstart
            text2hash(0) = j0
            text2hash(1) = j1
            text2hash(2) = j2
            text2hash(3) = j3
            
            For j = 0 To length - 1
                M = text2hash(j)
                calculate_hash_value
            Next j
                                        'MsgBox  (text2hash(0)  &  "  "  &  text2hash(1)  &  "  "  &
text2hash(2) & " " & text2hash(3) & " " & text2hash(4) & " " & " HASH VALUE=" & H)
        If H = result(i) Then
                                        same_hash_counter = same_hash_counter + 1
                                        If showresults = True Then MsgBox (Chr$(text2hash(0) - offset)
&  "  "  &  Chr$(text2hash(1)  -  offset)  &  "  "  &  Chr$(text2hash(2)  -  offset)  &  "  "  &
Chr$(text2hash(3) - offset) & " HASH VALUE=" & H)
                        End If
            Next
            Next
            Next
End If

If length = 5 Then
        For j0 = chr_start To chr_stop
        For j1 = chr_start To chr_stop
        For j2 = chr_start To chr_stop
```

```
            For j3 = chr_start To chr_stop
            j4 = buffer_chr

                H = Hstart
                text2hash(0) = j0
                text2hash(1) = j1
                text2hash(2) = j2
                text2hash(3) = j3
                text2hash(4) = j4
                For j = 0 To length - 1
                    M = text2hash(j)
                    calculate_hash_value
                Next j
                                    'MsgBox  (text2hash(0)  &  " "  &  text2hash(1)  &  " "  &
text2hash(2) & " " & text2hash(3) & " " & text2hash(4) & " " & " HASH VALUE=" & H)
        If H = result(i) Then
                                    same_hash_counter = same_hash_counter + 1
                                    If showresults = True Then MsgBox (Chr$(text2hash(0) - offset)
& " " & Chr$(text2hash(1) - offset) & " " & Chr$(text2hash(2) - offset) & " " &
Chr$(text2hash(3) - offset) & " " & Chr$(text2hash(4) - offset) & " HASH VALUE=" & H)
                        End If
            Next
            Next
            Next
            Next
End If

If length = 6 Then
            For j0 = chr_start To chr_stop
            For j1 = chr_start To chr_stop
            For j2 = chr_start To chr_stop
            For j3 = chr_start To chr_stop
            For j4 = chr_start To chr_stop
            j5 = buffer_chr

                H = Hstart
                text2hash(0) = j0
                text2hash(1) = j1
                text2hash(2) = j2
                text2hash(3) = j3
                text2hash(4) = j4
                text2hash(5) = j5
                For j = 0 To length - 1
                    M = text2hash(j)
                    calculate_hash_value
                Next j
                                    'MsgBox  (text2hash(0)  &  " "  &  text2hash(1)  &  " "  &
text2hash(2) & " " & text2hash(3) & " " & text2hash(4) & " " & " HASH VALUE=" & H)
        If H = result(i) Then
                                    same_hash_counter = same_hash_counter + 1
                                    If showresults = True Then MsgBox (Chr$(text2hash(0) - offset)
& " " & Chr$(text2hash(1) - offset) & " " & Chr$(text2hash(2) - offset) & " " &
Chr$(text2hash(3) - offset) & " " & Chr$(text2hash(4) - offset) & " " & Chr$(text2hash(5) -
offset) & " HASH VALUE=" & H)
                        End If
            Next
            Next
            Next
            Next
            Next
End If

If length = 7 Then
            For j0 = chr_start To chr_stop
            For j1 = chr_start To chr_stop
            For j2 = chr_start To chr_stop
            For j3 = chr_start To chr_stop
            For j4 = chr_start To chr_stop
            For j5 = chr_start To chr_stop
            j6 = buffer_chr

                H = Hstart
                text2hash(0) = j0
                text2hash(1) = j1
                text2hash(2) = j2
                text2hash(3) = j3
                text2hash(4) = j4
                text2hash(5) = j5
                text2hash(6) = j6
                For j = 0 To length - 1
                    M = text2hash(j)
                    calculate_hash_value
                Next j
```

```
                                'MsgBox   (text2hash(0)   &   "   "   &   text2hash(1)   &   "   "   &
text2hash(2) & " " & text2hash(3) & " " & text2hash(4) & " " & " HASH VALUE=" & H)
          If H = result(i) Then
                                      same_hash_counter = same_hash_counter + 1
                                      If showresults = True Then MsgBox (Chr$(text2hash(0) - offset)
& " " & Chr$(text2hash(1) - offset) & " " & Chr$(text2hash(2) - offset) & " " &
Chr$(text2hash(3) - offset) & " " & Chr$(text2hash(4) - offset) & " " & Chr$(text2hash(5) -
offset) & " " & Chr$(text2hash(6) - offset) & " HASH VALUE=" & H)
                                  End If
            Next
            Next
            Next
            Next
            Next
            Next
End If

If length = 8 Then
            For j0 = chr_start To chr_stop
            For j1 = chr_start To chr_stop
            For j2 = chr_start To chr_stop
            For j3 = chr_start To chr_stop
            For j4 = chr_start To chr_stop
            For j5 = chr_start To chr_stop
            For j6 = chr_start To chr_stop
            j7 = buffer_chr

                H = Hstart
                text2hash(0) = j0
                text2hash(1) = j1
                text2hash(2) = j2
                text2hash(3) = j3
                text2hash(4) = j4
                text2hash(5) = j5
                text2hash(6) = j6
                text2hash(7) = j7
                For j = 0 To length - 1
                    M = text2hash(j)
                    calculate_hash_value
                Next j
                                'MsgBox   (text2hash(0)   &   "   "   &   text2hash(1)   &   "   "   &
text2hash(2) & " " & text2hash(3) & " " & text2hash(4) & " " & " HASH VALUE=" & H)
          If H = result(i) Then
                                      same_hash_counter = same_hash_counter + 1
                                      If showresults = True Then MsgBox (Chr$(text2hash(0) - offset)
& " " & Chr$(text2hash(1) - offset) & " " & Chr$(text2hash(2) - offset) & " " &
Chr$(text2hash(3) - offset) & " " & Chr$(text2hash(4) - offset) & " " & Chr$(text2hash(5) -
offset) & " " & Chr$(text2hash(6) - offset) & " " & Chr$(text2hash(7) - offset) & " HASH
VALUE=" & H)
                                  End If
            Next
            Next
            Next
            Next
            Next
            Next
            Next
End If

End Sub
Private Sub hash_random_string_samestart()
'If length = 1 no sense

If length = 2 Then
            j0 = buffer_chr
            For j1 = chr_start To chr_stop

                H = Hstart
                text2hash(0) = j0
                text2hash(1) = j1

                For j = 0 To length - 1
                    M = text2hash(j)
                    calculate_hash_value
                Next j
                                'MsgBox   (text2hash(0)   &   "   "   &   text2hash(1)   &   "   "   &
text2hash(2) & " " & text2hash(3) & " " & text2hash(4) & " " & " HASH VALUE=" & H)
          If H = result(i) Then
                                      same_hash_counter = same_hash_counter + 1
                                      If showresults = True Then MsgBox (Chr$(text2hash(0) - offset)
& " " & Chr$(text2hash(1) - offset) & " HASH VALUE=" & H)
                                  End If
            Next
End If
```

```
If length = 3 Then
           j0 = buffer_chr
           For j1 = chr_start To chr_stop
           For j2 = chr_start To chr_stop

               H = Hstart
               text2hash(0) = j0
               text2hash(1) = j1
               text2hash(2) = j2

               For j = 0 To length - 1
                   M = text2hash(j)
                   calculate_hash_value
               Next j
                               'MsgBox (text2hash(0) & " " & text2hash(1) & " " &
text2hash(2) & " " & text2hash(3) & " " & text2hash(4) & " " & " HASH VALUE=" & H)
       If H = result(i) Then
                               same_hash_counter = same_hash_counter + 1
                               If showresults = True Then MsgBox (Chr$(text2hash(0) - offset)
& " " & Chr$(text2hash(1) - offset) & " " & Chr$(text2hash(2) - offset) & " " & " HASH VALUE="
& H)
                           End If
           Next
           Next
End If

If length = 4 Then
           j0 = buffer_chr
           For j1 = chr_start To chr_stop
           For j2 = chr_start To chr_stop
           For j3 = chr_start To chr_stop

               H = Hstart
               text2hash(0) = j0
               text2hash(1) = j1
               text2hash(2) = j2
               text2hash(3) = j3

               For j = 0 To length - 1
                   M = text2hash(j)
                   calculate_hash_value
               Next j
                               'MsgBox (text2hash(0) & " " & text2hash(1) & " " &
text2hash(2) & " " & text2hash(3) & " " & text2hash(4) & " " & " HASH VALUE=" & H)
       If H = result(i) Then
                               same_hash_counter = same_hash_counter + 1
                               If showresults = True Then MsgBox (Chr$(text2hash(0) - offset)
& " " & Chr$(text2hash(1) - offset) & " " & Chr$(text2hash(2) - offset) & " " &
Chr$(text2hash(3) - offset) & " HASH VALUE=" & H)
                           End If
           Next
           Next
           Next
End If

If length = 5 Then
           j0 = buffer_chr
           For j1 = chr_start To chr_stop
           For j2 = chr_start To chr_stop
           For j3 = chr_start To chr_stop
           For j4 = chr_start To chr_stop

               H = Hstart
               text2hash(0) = j0
               text2hash(1) = j1
               text2hash(2) = j2
               text2hash(3) = j3
               text2hash(4) = j4
               For j = 0 To length - 1
                   M = text2hash(j)
                   calculate_hash_value
               Next j
                               'MsgBox (text2hash(0) & " " & text2hash(1) & " " &
text2hash(2) & " " & text2hash(3) & " " & text2hash(4) & " " & " HASH VALUE=" & H)
       If H = result(i) Then
                               same_hash_counter = same_hash_counter + 1
                               If showresults = True Then MsgBox (Chr$(text2hash(0) - offset)
& " " & Chr$(text2hash(1) - offset) & " " & Chr$(text2hash(2) - offset) & " " &
Chr$(text2hash(3) - offset) & " " & Chr$(text2hash(4) - offset) & " HASH VALUE=" & H)
                           End If
           Next
```

```
                Next
                Next
                Next
End If

If length = 6 Then
            j0 = buffer_chr
            For j1 = chr_start To chr_stop
            For j2 = chr_start To chr_stop
            For j3 = chr_start To chr_stop
            For j4 = chr_start To chr_stop
            For j5 = chr_start To chr_stop

                H = Hstart
                text2hash(0) = j0
                text2hash(1) = j1
                text2hash(2) = j2
                text2hash(3) = j3
                text2hash(4) = j4
                text2hash(5) = j5
                For j = 0 To length - 1
                    M = text2hash(j)
                    calculate_hash_value
                Next j
                                'MsgBox (text2hash(0) & " " & text2hash(1) & " " &
text2hash(2) & " " & text2hash(3) & " " & text2hash(4) & " " & " HASH VALUE=" & H)
        If H = result(i) Then
                                same_hash_counter = same_hash_counter + 1
                                If showresults = True Then MsgBox (Chr$(text2hash(0) - offset)
& " " & Chr$(text2hash(1) - offset) & " " & Chr$(text2hash(2) - offset) & " " &
Chr$(text2hash(3) - offset) & " " & Chr$(text2hash(4) - offset) & " " & Chr$(text2hash(5) -
offset) & " HASH VALUE=" & H)
                        End If
            Next
            Next
            Next
            Next
            Next
End If

If length = 7 Then
            j0 = buffer_chr
            For j1 = chr_start To chr_stop
            For j2 = chr_start To chr_stop
            For j3 = chr_start To chr_stop
            For j4 = chr_start To chr_stop
            For j5 = chr_start To chr_stop
            For j6 = chr_start To chr_stop

                H = Hstart
                text2hash(0) = j0
                text2hash(1) = j1
                text2hash(2) = j2
                text2hash(3) = j3
                text2hash(4) = j4
                text2hash(5) = j5
                text2hash(6) = j6
                For j = 0 To length - 1
                    M = text2hash(j)
                    calculate_hash_value
                Next j
                                'MsgBox (text2hash(0) & " " & text2hash(1) & " " &
text2hash(2) & " " & text2hash(3) & " " & text2hash(4) & " " & " HASH VALUE=" & H)
        If H = result(i) Then
                                same_hash_counter = same_hash_counter + 1
                                If showresults = True Then MsgBox (Chr$(text2hash(0) - offset)
& " " & Chr$(text2hash(1) - offset) & " " & Chr$(text2hash(2) - offset) & " " &
Chr$(text2hash(3) - offset) & " " & Chr$(text2hash(4) - offset) & " " & Chr$(text2hash(5) -
offset) & " " & Chr$(text2hash(6) - offset) & " HASH VALUE=" & H)
                        End If
            Next
            Next
            Next
            Next
            Next
            Next
End If

If length = 8 Then
            j0 = buffer_chr
            For j1 = chr_start To chr_stop
            For j2 = chr_start To chr_stop
            For j3 = chr_start To chr_stop
            For j4 = chr_start To chr_stop
```

```
                        For j5 = chr_start To chr_stop
                        For j6 = chr_start To chr_stop
                        For j7 = chr_start To chr_stop
                        
                            H = Hstart
                            text2hash(0) = j0
                            text2hash(1) = j1
                            text2hash(2) = j2
                            text2hash(3) = j3
                            text2hash(4) = j4
                            text2hash(5) = j5
                            text2hash(6) = j6
                            text2hash(7) = j7
                            For j = 0 To length - 1
                                M = text2hash(j)
                                calculate_hash_value
                            Next j
                                            'MsgBox (text2hash(0) & " " & text2hash(1) & " " &
text2hash(2) & " " & text2hash(3) & " " & text2hash(4) & " " & " HASH VALUE=" & H)
            If H = result(i) Then
                                            same_hash_counter = same_hash_counter + 1
                                            If showresults = True Then MsgBox (Chr$(text2hash(0) - offset)
& " " & Chr$(text2hash(1) - offset) & " " & Chr$(text2hash(2) - offset) & " " &
Chr$(text2hash(3) - offset) & " " & Chr$(text2hash(4) - offset) & " " & Chr$(text2hash(5) -
offset) & " " & Chr$(text2hash(6) - offset) & " " & Chr$(text2hash(7) - offset) & " HASH
VALUE=" & H)
                                        End If
                 Next
                 Next
                 Next
                 Next
                 Next
                 Next
                 Next
End If

End Sub
Private Sub Command4_Click()
    For i = 0 To instruction_number
        If Check1(i).Value = False Then GoTo next_i 'do this only if the box next to the
instruction is checked
    same_hash_counter = 0

        ' Create random strings of the same length and same starting char and compare the hash
results
        length = Len(Text1(i).Text)
        ' Save the first character
        buffer_chr = Asc(Left$(Text1(i).Text, 1)) + offset
        ReDim text2hash(length)
        
            hash_random_string_samestart 'with same length
            same_hash_counter = same_hash_counter - 1 ' because it counted the original string
too
            occurence_p = ((same_hash_counter / (chr_stop - chr_start + 1) ^ length) * 100)
            Text4(i).Text = same_hash_counter & " /P=" & occurence_p & "%"

next_i:
    Next i
End Sub

Private Sub Command5_Click()
    For i = 0 To instruction_number
        If Check1(i).Value = False Then GoTo next_i 'do this only if the box next to the
instruction is checked
same_hash_counter = 0

        ' Create random strings of the same length and same starting char and compare the hash
results
        length = Len(Text1(i).Text)
        ' Save the first character
        buffer_chr = Asc(Right$(Text1(i).Text, 1)) + offset
        ReDim text2hash(length)
        
            hash_random_string_samestop 'with same length
            same_hash_counter = same_hash_counter - 1 ' because it counted the original string
too
            occurence_p = ((same_hash_counter / (chr_stop - chr_start + 1) ^ length) * 100)
            Text5(i).Text = same_hash_counter & " /P=" & occurence_p & "%"

next_i:
    Next i
End Sub
```

```
Private Sub Command6_Click()
    For i = 0 To instruction_number
        If Check1(i).Value = False Then GoTo next_i 'do this only if the box next to the instruction is checked
same_hash_counter = 0

        ' Create random strings of the same length and same starting char and compare the hash results
        length = Len(Text1(i).Text)
        ' Save the first character
        buffer_chr = Asc(Right$(Text1(i).Text, 1)) + offset
        buffer_chr2 = Asc(Mid$(Text1(i).Text, Len(Text1(i).Text) - 1, 1)) + offset
        ReDim text2hash(length)

            hash_random_string_same2Cstop 'with same length
            same_hash_counter = same_hash_counter - 1 ' because it counted the original string too
            occurence_p = ((same_hash_counter / (chr_stop - chr_start + 1) ^ length) * 100)
            Text8(i).Text = same_hash_counter & " /P=" & occurence_p & "%"

next_i:
    Next i
End Sub

Private Sub Form_Load()
    calculate_hash_value                    'called in order to get the hashname
    Form1.Caption = "evaluation of the iterative hash function:" & hashname
    M = 0
    H = 0
    initialize_values

End Sub

Private Sub Label5_DblClick()
'Select/deselect all
    If    select_instructions_switch    Then    select_instructions_switch    =    0    Else select_instructions_switch = 1
    For i = 0 To instruction_number            'hash all instructions
        Check1(i).Value = select_instructions_switch
    Next i
End Sub
```